\documentclass[aps,showpacs,nofootinbib,preprint,tightenlines]{revtex4}
\usepackage{psfig}
\usepackage{bm}

\begin{document}

\begin{titlepage}

\begin{flushright}
\vbox{
\begin{tabular}{l}
 Alberta Thy 18-02\\
 BNL-HET-02/25\\
 TPI-MINN-02/46\\
 UMN-TH-2119/02\\
 hep-ph/0212229
\end{tabular}
}
\end{flushright}

\vspace{0.6cm}

\title{%
Refinements in electroweak contributions \\ 
to the muon anomalous magnetic moment \vspace{4mm}
}

\author{Andrzej Czarnecki}
\affiliation{
Department of Physics, University of Alberta,
Edmonton, AB\   T6G 2J1, Canada}

\author{William J.~Marciano}
\affiliation{Physics Department, Brookhaven National Laboratory,
Upton, NY 11973, USA}

\affiliation{Institut f\"{u}r Theoretische Teilchenphysik,
Universit\"{a}t Karlsruhe,
D--76128 Karlsruhe, Germany}

\author{Arkady Vainshtein}
\affiliation{Theoretical Physics Institute, University of Minnesota,
116 Church St.\ SE, Minneapolis, MN 55455, USA \vspace{1.5cm}}

\begin{abstract}

\vspace{2mm}
\baselineskip 14pt

Effects of strong interactions on the two loop electroweak radiative
corrections to the muon anomalous magnetic moment,
$a_\mu=(g_\mu-2)/2$, are examined. Short-distance logs are shown to be
unaffected.  Computation of long-distance contributions is improved by
use of an effective field theory approach that preserves the chiral
properties of QCD and accounts for constraints from the operator
product expansion. Small, previously neglected, two loop
contributions, suppressed by a $1-4\sin^2\theta_W$ factor, are
computed and the complete three loop leading short-distance logs are
reevaluated.  These refinements lead to a reduction in uncertainties
and a slight shift in the total electroweak contribution to $a_\mu^{\rm
EW} = 154(1)(2)\times 10^{-11}$ where the first error corresponds to
hadronic uncertainties and the second is primarily due to the allowed
Higgs  mass range.
\end{abstract}
 

\maketitle
\thispagestyle{empty}
\end{titlepage}

\tableofcontents
\newpage

\section{Introduction}
   Recently, experiment E821 at Brookhaven National Laboratory
achieved an order of magnitude improvement (relative to the classic
CERN experiments) in the determination of the muon anomalous magnetic
moment, $a_\mu = (g_\mu-2)/2\,$.  The new world-average value for that
fundamental quantity is~\cite{Bennett:2002jb}
\begin{eqnarray}
        a_\mu = 116\,592\,030(80) \times 10^{-11}\,.
\label{e821}
\end{eqnarray}
Additional data currently being analyzed should further reduce the
uncertainty.

At the present level of precision, comparison with the theoretical
prediction for $a_\mu$ from the Standard Model requires knowledge of
hadronic vacuum polarization effects with accuracy of one percent. The
most recent dispersion integral 
analysis \cite{Davier:2002dy} (see also \cite{Hagiwara:2002ma}) based on
data from electron-positron annihilation into hadrons and hadronic
$\tau$ decays demonstrates that the issue is unsettled: $e^+e^-$
annihilation leads to a prediction lower by about $3\,\sigma$ than the
value (\ref{e821}) while the prediction based on $\tau$ data is lower
only by $1.3\,\sigma$. Moreover, the vector spectral functions derived
from $e^+e^-$ annihilation and from $\tau$ decays differ significantly for
energies beyond the $\rho$ resonance peak (by more than
10\% in some regions).  It seems very
difficult to explain such a large difference by isospin breaking
effects.  Thus, it appears that the data from $e^+e^-$
annihilation and from $\tau$ decays are incompatible: so, no
conclusion can be derived yet about a deviation from the Standard
Model prediction. 

Since a real deviation from theory would signal the presence of
``new physics'', with supersymmetry the leading candidate, it is
extremely important that all such hadronic uncertainties be thoroughly
scrutinized and eliminated as much as possible before implications are
drawn.   Toward that end, new $e^+e^-$ and $\tau$ data from Frascati
and  the $B$ factories will hopefully help to resolve this puzzling
difference.

     Beyond the leading hadronic vacuum polarization effects, strong
interaction uncertainties also enter $a_\mu$ via higher orders
that involve quark loops. Quark loops appear in light-by-light
scattering contributing in three loops  as well as in two-loop
electroweak corrections. The latter are the subject of this paper 
although the hadronic uncertainties there are certainly much smaller than
those induced by  light-by-light scattering.

At the two loop electroweak level, hadronic uncertainties arise from
two types of diagrams, quark triangle diagrams related to the anomaly
and hadronic photon-$Z$ mixing. The first category has been previously
studied in a free quark approximation and the more general operator
product expansion. Although phenomenologically both approaches produce
very close numbers they differ: particularly with regard to their
explicit short distance dependence, i.e.\ $\log m_Z$ terms.  Here, we
show that this difference is due to an incomplete operator product
analysis in the second approach.  When corrected, unambiguous
short-distance contributions result.  We also take this opportunity to
update the long distance and total electroweak contributions.

In the case of photon-$Z$ mixing, its two loop contribution to
$a_\mu$ is suppressed by a factor $1\!-\!4\sin^2 \theta_W \sim 0.1$; so,
it is not as important. It can be evaluated either in the free quark
approximation (sufficient for logarithmic accuracy) or via a
dispersion  relation using data from $e^+e^-$ annihilation into
hadrons. The difference is shown to be numerically insignificant.

Finally, having clarified the leading short-distance behavior of the
two loop electroweak radiative corrections to $a_\mu$, we can use the
renormalization group to estimate higher order leading-log
contributions which, due to an interesting  cancellation, 
turn out to be very small.

In the end, our analysis leads to a new, not very different, but more
precise and better founded prediction for the electroweak
contributions to $a_\mu$.

\section{Electroweak contributions to $\bm{a_\mu}$}
\label{sec:EW}

\begin{figure}[ht]
\hspace*{0mm}
\begin{minipage}{16.cm}
\begin{tabular}{c@{\hspace{16mm}}c@{\hspace{16mm}}c}
\psfig{figure=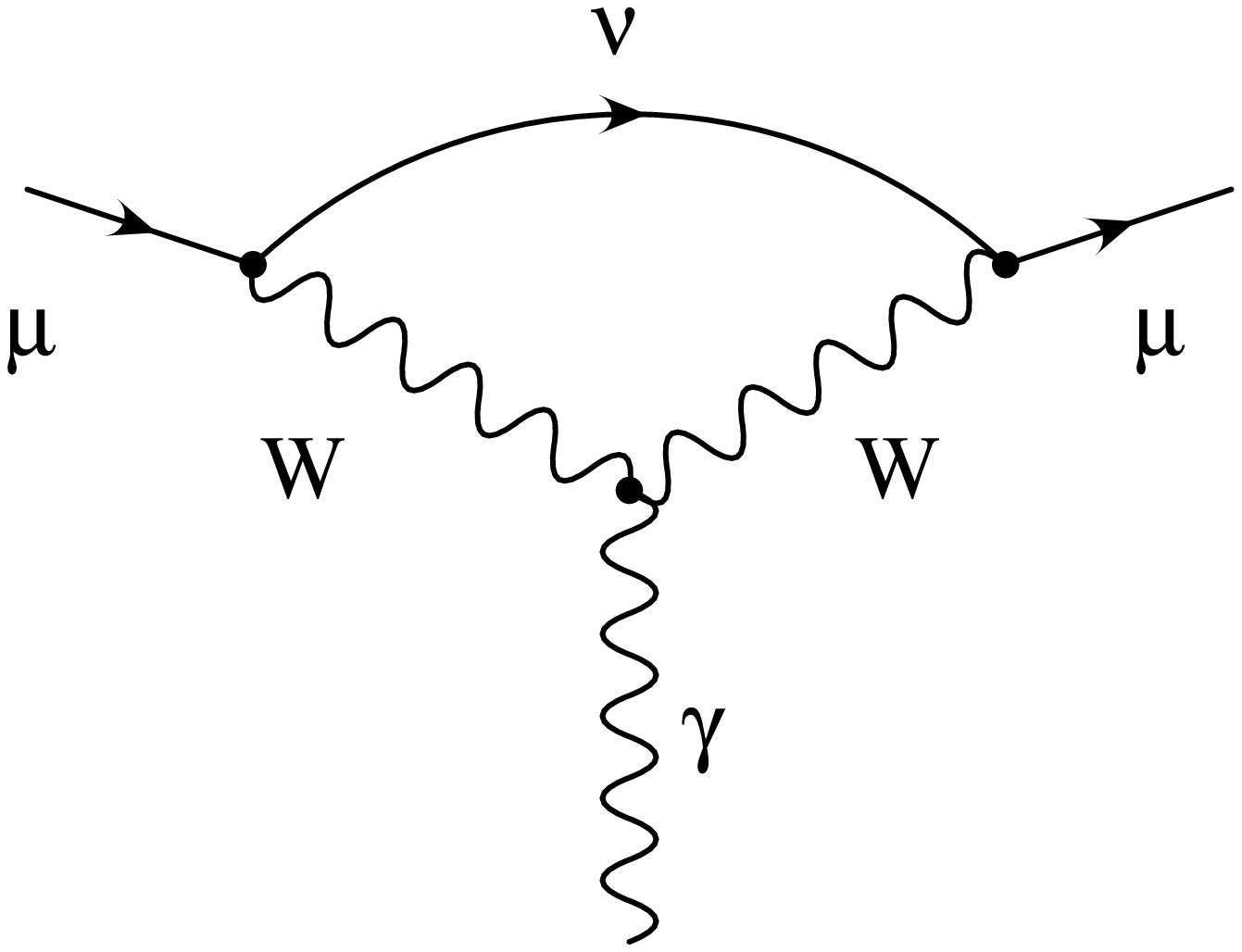,width=40mm}
&
\psfig{figure=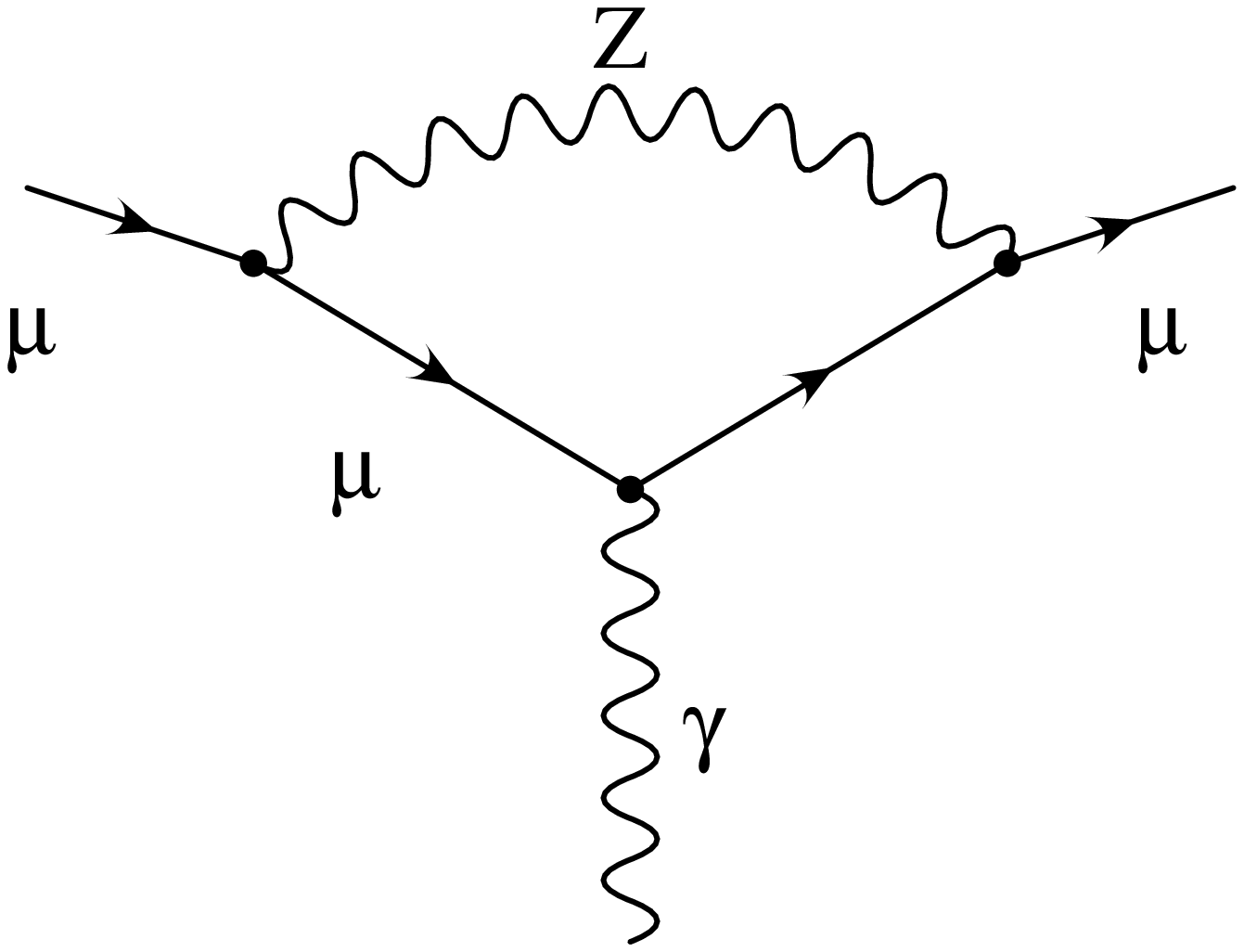,width=40mm}
&
\psfig{figure=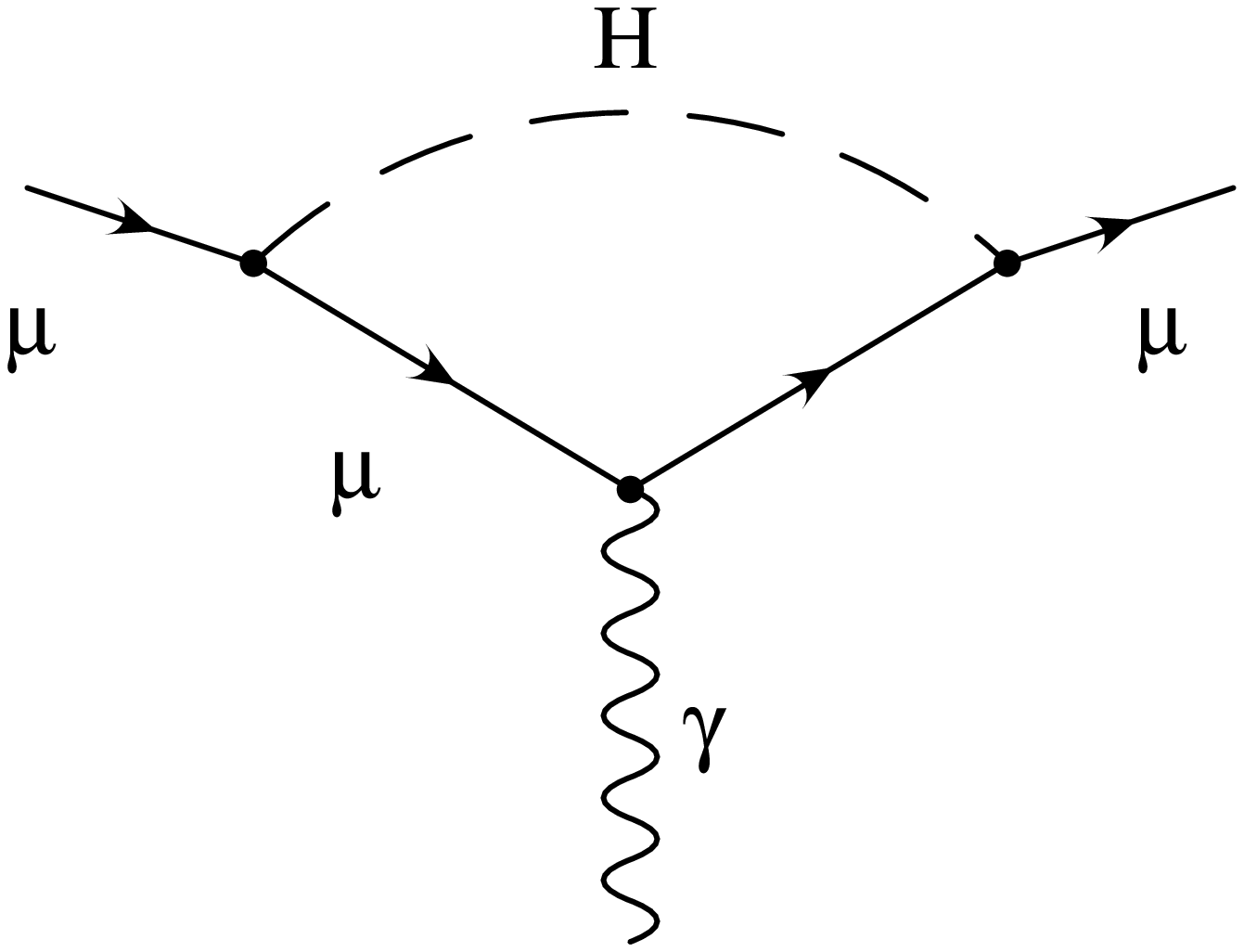,width=40mm}
\\[1mm]
\end{tabular}
\end{minipage}
\caption{\sf \small One-loop electroweak contributions to $a_\mu$.} 
\label{fig:oneloop}
\end{figure}

In the Standard Model (SM) the one-loop electroweak contributions to
$a_\mu$, illustrated in Fig.\,\ref{fig:oneloop}, were computed about 30
years ago \cite{fls72,Jackiw72,ACM72,Bars72,Bardeen72}.  They have the
relatively simple form
\begin{eqnarray}
a_\mu^{\rm EW}(\mbox{1-loop}) = {5\,G_\mu\, m_\mu^2 \over 24 \sqrt{2}\pi^2}
\left[ 1 + {1\over 5}\, (1-4\sin^2\theta_W)^2
 +{\cal O}\left({{m_\mu^2 \over m_{W,H}^2}}\right)\right] 
\label{eq10}
\end{eqnarray}
where $G_\mu = 1.16637(1)\times 10^{-5} \, {\rm GeV}^{-2}$
is the Fermi constant obtained from the muon lifetime and $\theta_W$
is the weak mixing angle.  

For $\sin^2\theta_W$ we employ the on-shell renormalized
definition
\begin{eqnarray}
\sin^2\theta_W \equiv s_W^2 = 1-{m_W^2\over m_Z^2}
\label{eq12}
\end{eqnarray}
where $m_Z = 91.1875(21) \, {\rm GeV}$
and the $W$ mass is correlated with the Higgs scalar mass, $m_H$, via
loop corrections (for $m_t=174.3$ GeV)  \cite{Ferroglia:2002rg}
\begin{eqnarray}
m_W = \left( 80.373-0.05719\ln{m_H \over 150 \, {\rm GeV}}
 -0.00898\ln^2 {m_H \over 150 \, {\rm GeV}} \right)  \, {\rm GeV}
\label{eq14}
\end{eqnarray}
For $m_H=150$ GeV, the central value employed in this paper, we must
use $m_W=80.373$ GeV (rather than the direct experimental value, $m_W
= 80.451(33)$ GeV, which corresponds to a very
small $m_H$), for SM loop consistency.  That implies
\begin{eqnarray}
s_W^2 =0.2231
\label{eq15}
\end{eqnarray}
and 
\begin{eqnarray}
a_\mu^{\rm EW}(\mbox{1-loop}) = 194.8\times 10^{-11}.
\label{eq16}
\end{eqnarray}

The calculation of two loop electroweak contributions to $a_\mu^{\rm
EW}$ was more recent and considerably more involved.  It started with
the observation by Kukhto et al.\,\cite{KKSS} that some two-loop
electroweak diagrams were enhanced by large logs of the form $\ln
(m_Z/m_\mu)$. Those authors carried out detailed calculations for a
number of such enhanced diagrams. They did not account, however, for
closed quark loops.  At about the same time, in
Ref.~\cite{D'Hoker:1992bv} it was shown that for superheavy fermions,
like the top quark, logarithms of their mass appear in corrections to
magnetic moments due to triangle anomaly diagrams. Detailed studies of
all closed quark loops were included in the calculation of $a^{\rm
EW}_\mu$ in Refs.~\cite{Peris:1995bb,CKM95}.  Finally, in
Ref.~\cite{CKM96}, the two-loop calculation of all logs as well
as constant terms was completed.
               
   Two loop corrections to $a_\mu^{\rm EW}$ naturally divide into
leading logs (LL), i.e. terms enhanced by a factor of $\ln(m_Z/m_f)$
where $m_f$ is a fermion mass scale much smaller than $ m_Z$, and
everything else, which we call non-leading logs (NLL).  The 2-loop
leading logs are (see Eq.\,(\ref{total2}) below)
\begin{eqnarray}
a_\mu^{\rm EW} (\mbox{2-loop})_{LL} &=& {5G_\mu m_\mu^2 \over 24
\sqrt{2}\pi^2}
\cdot {\alpha\over \pi}\left\{
-{43\over 3} \left[1+{31\over 215} (1-4s_W^2)^2\right]\ln{m_Z\over
m_\mu}
\right.
\nonumber \\
&+&
\left.
{36\over 5} \sum_{f\in F} N_f Q_f \left[
I^3_f\, Q_f -{2\over 27} \left( I^3_f - 2Q_f s_W^2\right)
\left( 1-4s_W^2\right)\right] \ln{ m_Z\over m_f} \right\},
\label{eq17}
\end{eqnarray}
where $\alpha \simeq  1/137.036$,  $N_f=3$ for quarks and 1 for
leptons,  $I^3_f$ is the 3rd
component of weak isospin and $Q_f$ is electric charge. Electron and
muon loops as well as non-fermionic loops produce the $\ln(m_Z/
m_\mu)$ terms in this expression (the first line) while the sum runs    
over $F=\tau,u,d,s,c,b$. 
The logarithm $\ln(m_Z/m_f)$ in the sum implies that the fermion mass 
$m_f$ is larger than $m_\mu$. For the light 
quarks, such as $u$ and $d$, whose current masses are very
small, $m_f$ has a meaning of some effective hadronic mass scale.

\begin{figure}[ht]
\hspace*{0mm}
\begin{minipage}{16.cm}
\psfig{figure=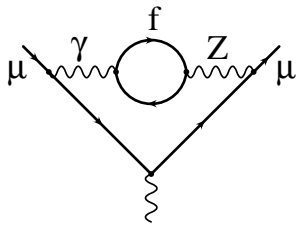,width=40mm}
\end{minipage}
\caption{\sf \small Contribution to $a^{\rm EW}_\mu$
from the $\gamma$-$Z$ mixing induced by a fermion $f$.}
\label{fig:gZmix}
\end{figure}

In Eq.\,(\ref{eq17}) we have retained for completeness
small contributions from the $\gamma$-$Z$ mixing diagrams in
Fig.\,\ref{fig:gZmix} which were previously excluded because they are
suppressed by $(1-4 s_W^2)$ for quarks and $(1-4 s_W^2)^2$ for
leptons. 

The more important terms in Eq.\,(\ref{eq17}), those not suppressed by 
$(1-4 s_W^2)$, were checked by Degrassi and
Giudice \cite{Degrassi:1998es}.
They used knowledge of well studied QCD corrections to $b\to s\gamma$
decay and translated them into into QED
corrections to $a_\mu^{\rm EW}$ via appropriate coupling changes.
The only place that they erred was for the small 
$-{2\over 27} \left( I^3_f - 2Q_f s_W^2\right)
\left( 1-4s_W^2\right) \ln{ m_Z\over m_f}$ terms coming from quark
loops.  Technically the difference is due to  $\gamma$-$Z$ mixing 
(Fig.\,\ref{fig:gZmix}) which is proportional to $Q_f Q_\mu$ (electric
charges of the loop fermion and the muon): in
\cite{Degrassi:1998es} it was given as $Q_f^2$.
\begin{figure}[ht]
\hspace*{0mm}
\begin{minipage}{16.cm}
\psfig{figure=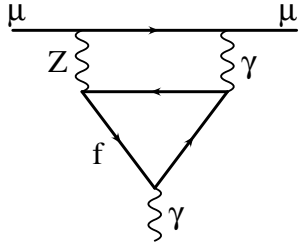,width=40mm}
\end{minipage}
\caption{\sf \small Effective $Z\gamma\gamma^*$ coupling induced by a
fermion triangle, contributing to $a^{\rm EW}_\mu$.}
\label{fig:triangle}
\end{figure}

In addition to leading logs, the NLL 2-loop contributions have also
been computed \cite{CKM96}.  They depend on known constants, the top
quark mass (here taken to be 174.3 GeV),  $\ln(m_Z/m_W)$ terms, 
and the as yet unknown Higgs
mass, $m_H$.  For $m_H\simeq 150$ GeV, those corrections are
numerically given by \cite{CKM96}
\begin{eqnarray}
a_\mu^{\rm EW} (\mbox{2-loop})_{NLL} = -6.0\pm 1.8 \times 10^{-11},
\label{eq18}
\end{eqnarray} 
where the error allows $m_H$ to range from $m_H\simeq 114 $ GeV
(its experimental lower bound) to about 250 GeV. We have included in
Eq.\,(\ref{eq18}) small NLL contributions, $-0.2\times 10^{-11}$,
proportional to $(1-4s_W^2)\,m_t^2/m_W^2$ induced by the
renormalization of the weak mixing angle.

When evaluating Eq.\,(\ref{eq17}), one is confronted by the presence of
light $u$, $d$, and $s$ quark masses in the logarithms.  They were
used to crudely regulate long distance loop contributions in
Figs.\,\ref{fig:gZmix} and \ref{fig:triangle}, where QCD effects were
ignored \cite{CKM95}. For $m_{u,d}=300$ MeV, $m_s=500$~MeV, $m_c=1.5$
GeV, and $m_b=4.5$ GeV, one finds
\begin{eqnarray}
a_\mu^{\rm EW} (\mbox{2-loop})_{LL} = -(36.7\pm 2) \times 10^{-11}\,,
\label{eq19}
\end{eqnarray}
where the error is meant to roughly
reflect low-momentum hadronic loop
uncertainties for the  $u$, $d$, and $s$ quarks in
Fig.\,\ref{fig:triangle}.  Together, Eqs.\,(\ref{eq18}) and (\ref{eq19})
provide the two-loop total electroweak correction
$a_\mu^{\rm EW} (\mbox{2-loop}) = -42.7(2)(1.8) \times   10^{-11}$, 
which together with Eq.\,(\ref{eq16}) leads to the generally
quoted Standard Model prediction \cite{Czarnecki:2001pv},
\begin{eqnarray}
a_\mu^{\rm EW} = 152(4)  \times   10^{-11},
\label{eq20}
\end{eqnarray}
where the error of $\pm 4\times 10^{-11}$ is meant to reflect the total
uncertainties coming from hadronic loop effects, the unknown Higgs
mass and uncalculated higher order (three-loop) contributions.

Refinements in the above analysis are possible on two fronts:
improvement in the low-momentum contribution of hadronic loops and an
estimation of the leading three-loop electroweak contribution (which is
part of the overall uncertainty).  The Higgs mass uncertainty will be
overcome with its discovery.

The easiest hadronic loop improvement can be made in the quark
contributions to $\gamma$-$Z$ mixing pictured in Fig.\,\ref{fig:gZmix}.
Those effects, embodied in the last part of the bracketed terms in
Eq.\,(\ref{eq17}) can be obtained via a dispersion relation using
$\sigma(e^+e^-\to {\rm hadrons})$ data.  Such an analysis has been
performed for various low energy processes.  It effectively leads to
the replacement \cite{Czarnecki:1996fw}
\begin{eqnarray}
-{2\over 3} \sum_{q=u,d,s,c,b} N_q
\left( I^3_{q}\, Q_q - 2\,Q_q^2\,s_W^2\right)
\ln{m_Z\over m_q} \to -6.88\pm 0.50
\label{eq21}
\end{eqnarray}
which is somewhat larger than the value $(-5.95)$ obtained in 
logarithmic approximation with
constituent $u$, $d$, and $s$ quark masses.  (It suggests that
smaller quark masses might be more appropriate.)
Because those $\gamma$-$Z$ mixing effects are suppressed by $1\!-\!4s_W^2$,
the shift in $a_\mu^{\rm EW}$ from this modification is tiny,
$-0.02\times 10^{-11}\,$, and can be safely neglected.

Low momentum loop effects in the light quark triangle diagrams of
Fig.\,\ref{fig:triangle} are more important. It is clear that the
use of effective quark mass as an infrared cut off is in contradiction
with the chiral properties of QCD in the case of light quarks. Indeed, in the
chiral limit the infrared singularity in the quark triangle does not
go away: it matches the Goldstone pole in hadronic
terms.  This refers 
to the anomalous part of the triangle, i.e.\ to the longitudinal part of
the axial current.    

The issue of how to properly treat light quark triangle diagrams, was
addressed originally in a study by Peris, Perrotet and de Rafael
\cite{Peris:1995bb} within a low-energy effective field theory
approach for $\pi$, $\eta$, and $\eta'$ mesons.  More recently,
Knecht, Peris, Perrotet and de Rafael \cite{Knecht:2002hr} have
reexamined the issue using an operator product expansion (OPE) and
Ward identities as guidance.  We find their approach to the anomaly
related longitudinal part of quark triangles to be a valid improvement
over the naive constituent quark mass cutoff of Eq.\,(\ref{eq17}).

Unfortunately, their rather sophisticated OPE analysis failed to
properly address the short-distance behavior of nonanomalous, i.e.\
transversal, part of the light quark triangles in
Fig.\,\ref{fig:triangle}. In particular, they do not reproduce the
complete $\ln m_Z$ terms in Eq.\,(\ref{eq17}).  That difference was
attributed to QCD damping effects in \cite{Knecht:2002hr} which were
claimed to eliminate the nonanomalous $\ln m_Z$ light quark contributions.
However, in our opinion, it points to a shortcoming in their analysis.

In section \ref{sec:had}, we address in some detail what modifications
to the study in \cite{Knecht:2002hr} are required to restore the
proper short-distance behavior.  We then employ the effective field
theory approach to improve the evaluation of light quark diagrams in
Fig.\,\ref{fig:triangle}, thus refining their contribution to
$a_\mu^{\rm EW}$.  

Having verified the short-distance behavior of
Eq.\,(\ref{eq12}) we are also in a position to evaluate higher order
leading logs via the renormalization group.  That analysis is carried
out in section \ref{sec:logs}, where the leading log 3-loop
contributions to $a_\mu^{\rm EW}$ are determined.  Such a study was
previously undertaken by Degrassi and Giudice \cite{Degrassi:1998es}.
Although we find small differences with their analysis, in the end we
also obtain a very small leading log 3-loop contribution to
$a_\mu^{\rm EW}$.  In fact, the result is consistent with zero, to our
level of accuracy, due to an interesting cancellation between
anomalous dimensions and beta function effects.  

Finally, in section
\ref{sec:five}, we give a refined determination of $a_\mu^{\rm EW}$
for which the errors are reduced and the central value is slightly
shifted due to improvements in our analysis.

\section{Hadronic effects in quark triangles}
\label{sec:had}

An interesting subset of the two-loop contributions to $a_\mu^{\rm
EW}$ are those containing fermionic triangles of quarks and leptons, 
see Fig.\,\ref{fig:triangle}. The internal triangles define the
one-loop $Z^*\gamma\gamma^*$ amplitude where the $Z$ and one photon are
virtual while the other photon is real and soft. Those same triangles produce
the well-known anomaly part of the $Z$ boson axial-current.  For
cancellation of the anomaly, one needs to sum over all fermions in a
given generation.

   In the case of $a_\mu^{\rm EW}$, the cancellation between quarks
and leptons in Fig.\,\ref{fig:triangle} is not complete, because of
their different masses and interactions.  In this section we give a
detailed analysis of the general structure of the $Z^*\gamma\gamma^*$
amplitude, paying particular attention to the effect of strong
interactions on the quark diagrams. Perturbative QCD corrections to
short-distance logs are shown to vanish and have an overall negligible
effect for heavy quark diagrams.  In the case of light quarks, an
operator product expansion and effective field theory approach are
used to improve the evaluation of long-distance QCD effects.  These
refinements lead to an update of the fermionic triangle loop
contributions to $a_\mu^{\rm EW}$.

\subsection{Structure of the \bm{$Z^*\gamma \gamma ^*$} interaction}
\label{stru}

We begin our general analysis by
introducing some definitions. The interaction Lagrangian for the
electromagnetic and $Z$-boson fields, $A_\mu $ and $Z_\nu $, is
\begin{eqnarray}
{\cal L}_{\rm int} = e A^\mu\,j_\mu  
 -\frac{g}{4\cos \theta_W }\, Z^\nu j^5_\nu\;,\qquad 
j_\mu = \sum_f Q_f \bar f \gamma _\mu f \,,\quad
j^5_\nu =\sum_f 2I^3_{f} \bar f \gamma _\nu \gamma _5 f
\,,
\end{eqnarray}
where $Q_f$ and $I^3_{f}$ are the electric charge and the third
component of weak isospin and we retain only the axial part of the
$Z$-boson current.  (The weak vector current contribution in
Fig.\,\ref{fig:triangle} vanishes by Furry's theorem.)

The $Z^*\gamma \gamma^*$ amplitude
$T_{\mu \nu }$ is defined as
\begin{equation}
T_{\mu \nu }=i\int {\rm d}^4 x\, {\rm e}^{iqx}\,\langle 0|\,T\{j_\mu
(x)\, j^5_\nu (0)\}|\,\gamma (k)\rangle\,.
\label{Tmunu}
\end{equation}
That is equivalent to 
\begin{equation}
T_{\mu \nu }=e\, e^\gamma\,T_{\mu\gamma\nu}\,, \qquad 
T_{\mu\gamma\nu}=- \int {\rm d}^4 x\,{\rm d}^4 y\, 
{\rm e}^{iqx-iky}\,\langle 0|\,T\{j_\mu
(x)\, j_\gamma(y)\,j^5_\nu (0)\}|0\rangle\,,
\label{Tmunu1}
\end{equation}
where $e^\gamma $ is the photon polarization vector.  We consider the
limit of small photon momentum $k$. The expansion of $T_{\mu \nu }$ 
in $k$ starts with linear terms and we
neglect quadratic and higher powers of $k$. In this approximation
there are two Lorentz structures for $T_{\mu \nu }$ consistent with
electromagnetic current conservation,
\begin{eqnarray}
\label{invfun}
T_{\mu \nu }&=& -\frac{ie}{4\pi^2}\left[w_T(q^2) 
\left(-q^2 \tilde f_{\mu \nu
}+q_\mu q^\sigma
\tilde f_{\sigma  \nu } - q_\nu q^\sigma \tilde f_{\sigma  \mu }\right)
+w_L(q^2)\, q_\nu q^\sigma \tilde f_{\sigma  \mu }\right],\\
\tilde f_{\mu \nu }&=&\frac{1}{2}\, \epsilon _{\mu \nu 
\gamma
\delta }  f^{\gamma \delta }\,,\qquad  f_{\mu \nu }=k_\mu e_\nu -k_\nu
e_\mu
\,.\nonumber
\end{eqnarray}
The first structure is transversal with respect to the axial current
index $\nu $, the second is longitudinal.

The contribution of $Z^*\gamma \gamma^*$ to the muon anomalous
magnetic moment $a^{\rm EW}_\mu$ in the unitary gauge where the
$Z$-propagator is $i (-g_{\mu \nu} +q_\mu q_\nu/m_Z^2)/(q^2-m_Z^2)$
can be written in terms of $w_{T,L}(q^2)$ as
\begin{eqnarray}
\label{amug}
\Delta a^{\rm EW}_\mu &=& \frac{\alpha}{\pi }\,2\sqrt{2}\,  G_\mu 
\,m_\mu^2\,i\!\int\!\frac{{\rm
d}^4 q}{(2\pi )^4}\frac{1}{q^2+2qp}
\nonumber\\[1mm]
&&\times
\left[\frac{1}{3}\left(1+\frac{2(qp)^2}{q^2 m_\mu^2}\right)
\left(w_L
-\frac{m_Z^2}{m_Z^2-q^2}\,w_T\right)
+\,\frac{m_Z^2}{m_Z^2-q^2}\,w_T\right]
.
\end{eqnarray}
Here $p$ is the four-momentum of the external muon.
For logarithmic estimates, a much simpler expression is sufficient,
\begin{equation}
\label{amu}
\Delta a^{\rm EW}_\mu
=\frac{\alpha}{\pi }\,\frac{G_\mu\,m_\mu ^2}{8\pi^2\sqrt{2}}
\int_{m_\mu^2}^\infty {\rm d}Q^2 \left(w_L+\frac{m_Z^2}{m_Z^2+Q^2}\,w_T
\right),
\end{equation}
where $Q^2=-q^2$.  Moreover, the same expression with the lower limit of
integration set to zero works with a power accuracy (in
$m_\mu^2/m_f^2$) in the case of a heavy fermion in the loop, $m_f\gg
m_\mu$.

The one-loop results for $Z^*\gamma \gamma ^*$ can be taken from the
classic papers by Adler~\cite{Adler:1969gk} and
Rosenberg~\cite{Rosenberg:1963pp}. In Ref.~\cite{KKSS} they were
considerably simplified in the limit of small external photon
momentum. One then finds the following one-loop expressions for invariant
functions $w_{L,T}$,
\begin{equation}
\label{wlt}
w^{\rm 1-loop}_L=2\,w^{\rm 1-loop}_T=\sum_f 4\,I^3_f \,N_f\,Q_f^2\int_0^1
\frac{{\rm d}\alpha\, \alpha (1-\alpha )}{\alpha (1-\alpha )Q^2+m_f^2}\,,
\end{equation}
where the factor $N_f$ accounts for colors  in the case of quarks.

We also independently calculated $T_{\mu \nu }$ using Schwinger
operator methods for the fermionic loop.  It can be presented as the
polarization operator describing the mixing of two currents, $j_\mu $
and $j^5_\nu $ but with the fermion propagators taken in the external
field with the constant field strength.
In the fixed point gauge $x^\mu  A_\mu =0$ this propagator has the
form \cite{SVNet}
\begin{equation}
\label{prop}
S(p)=\frac{1}{/\!\!\!p-m}+\frac{1}{(p^2-m^2)^2}\,eQ\, \tilde F_{\rho \delta
}\left(p^\rho
\gamma^\delta - \frac{i}{2}\,m\,\sigma^{\rho \delta }\right)\gamma _5
+{\cal O}(F^2)
\,.
\end{equation}
Then straightforward calculations lead to the above expressions
(\ref{wlt}) for the invariant functions $w_{T,L}(q^2)\,$.

The corresponding two-loop contributions
to $a^{\rm EW}_\mu$ were calculated in Refs.~\cite{Peris:1995bb,CKM95}.
According to Eq.\,(\ref{amug}) one needs  to integrate using the
$w_{T,L}$ given in Eq.\,(\ref{wlt}). Let us consider the part
$w_{T,L}[f]$ which is due to the loop of
a given fermion $f$ with the mass $m_f$  at the range of  
$Q^2\gg m_f^2\,$.
The asymptotic behavior of  $w_{T,L}[f]$ at large $Q^2$ is
\begin{equation}
\label{msquare}
w^{\rm 1-loop}_L[f]=2\,w^{\rm 1-loop}_T[f]=
4\,I^3_f \,N_f\,Q_f^2\left[\frac{1}{Q^2}-\frac{2\,m_f^2}{Q^4}\,
\ln\frac{Q^2}{m_f^2}+{\cal O}\left(\frac{1}{Q^6}\right)\right].
\end{equation}
At large $Q^2$ we can use the simpler form (\ref{amu}) for 
integration. It is clear then that for the individual fermion loop the
integral $\int {\rm d} Q^2 w_L$ is divergent. This reflects the fact
that the theory is inconsistent unless the condition of anomaly
cancellation between leptons and quarks is fulfilled. This
condition has the form
\begin{equation}
\sum_f I^3_f \,N_f\,Q_f^2=0
\label{anomcan}
\end{equation}
within every generation. It means that at $Q^2\gg m_f^2$ the leading terms
$1/Q^2$ cancel out after summing over fermions and $w_{L}\sim (\ln
Q^2)/Q^4$  implying 
convergence for $a^{\rm EW}_\mu\,$.

Note the difference between the $w_L$ and $w_T$ parts in the integral
(\ref{amu}). The first one does not depend on $m_Z$ at all while for the
second we have a cut off factor $1/(Q^2+m_Z^2)$. Therefore, the $w_T$
part is never divergent, instead individual fermion loops produce
$\log (m_Z/m_f)$ terms in $a^{\rm EW}_\mu$ when $m_f\ll m_Z$. On the
other hand, the one-loop relation $w_T=w_L/2$ means the anomaly
cancellation condition (\ref{anomcan}) leads to cancellation of the
leading $1/Q^2$ terms in $w_T$ as well.  It results in absence of
$\log m_Z$ terms when $m_f\ll m_Z$ for all fermions in the given
generation.

\subsection{Hadronic corrections for quark triangles}
\label{hadroncor}

How good is the one-loop approximation for $w_L$ and $w_T$? This
question pertains to strong interaction effects for quark loops. As
characterized in \cite{Peris:1995bb} this issue brings about a new
class of hadronic contributions to the muon anomalous magnetic moment.

Let us first discuss perturbative corrections to the $Z\gamma\gamma*$
amplitude at $Q\gg m_q$ due to gluon exchange in quark loops.
The longitudinal function $w_L$
is protected against these corrections by a nonrenormalization theorem
\cite{Adler:er}
for the anomaly. What about the transversal function $w_T$?  If the
$\alpha_s$ corrections were present for $w_T$ then after summing over the
fermion generation, the leading term would become
$\alpha_s(Q)/Q^2\sim 1/(Q^2\log Q/\Lambda_{\rm QCD})$. According to
Eq.\,(\ref{amu}) this leads to terms in $a^{\rm EW}_\mu$ which are
parametrically enhanced by $\log(\log m_Z/\Lambda_{\rm QCD})$.

It turns out, however, that the $\alpha_s$ corrections in $w_T$ are
also absent at $ Q \gg m_q$ due to the new nonrenormalization theorem
proved in Ref.~\cite{WT}. The proof, stimulated by the present study,
is based on preservation of the relation $w_T=w_L/2$ beyond the one
loop, i.e.\ in the presence of QCD interactions. This relation holds
only for the specific kinematics we consider in which the external
photon momentum is vanishing.

Our above discussion means that $\alpha_s$ corrections
are absent for both $w_L$ and $w_T$ in the chiral limit $m_q=0$. 
When quarks are heavy the $\alpha_s$ corrections can appear 
but with a suppression factor $m_q^2/Q^2$ at $Q\gg m_q$.
In application to $a^{\rm EW}_\mu$ it implies that perturbative $\alpha_s$
corrections are absent for light quarks. For heavy quarks, the
logarithmic terms which are due to  $Q\gg m_q$ are not renormalized
but non-logarithmic terms regulated by $\alpha_s(m_q)$ could appear
due to the range of momenta $Q\sim m_q$.

Next comes the question of nonperturbative corrections.  
For the heavy quarks these corrections, given by some power of
$\Lambda_{\rm QCD}^2/m_q^2\,$, are small. As  discussed above,  
the perturbative strong interaction corrections governed 
by $\alpha_s(m_q)$ are also small for heavy quarks.
In particular, this argument is applicable to the third
generation, $\tau $, $b$ and $t$ loops, so the free quark 
computational results
obtained in Refs.~\cite{Peris:1995bb,CKM95,Knecht:2002hr} are very
much under theoretical control.

The first and the second generations contain light quarks $u,\ d,\ s$
for which the momentum range of $Q$ spans the hadronic scale
$\Lambda_{\rm QCD}$ where nonperturbative effects are ${\cal
O}\left({100\%}\right)$ 
and give unsuppressed contributions to $a^{\rm EW}_\mu $.  This
problem has been addressed in the literature and two approaches were
suggested. In Ref.~\cite{CKM95} effective quark masses for light
quarks in one-loop expressions were introduced as a simple way to
account for strong interactions.  This mass plays the role of an
infrared cutoff in the integral over $Q$.  A more realistic approach
to the relevant hadronic dynamics was worked out in
Refs.~\cite{Peris:1995bb,Knecht:2002hr}. Unfortunately, some
conceptual mistakes in applying the OPE (we are going to comment on
them in more detail in Sec.~\ref{sec:OPE}) led to incorrect
results. This is immediately obvious in the ultraviolet sensitivity of
the results in Ref.~\cite{Knecht:2002hr}: the dependence on $\ln m_Z$ is
not suppressed for the first and the second generations where all the
masses are much less than $m_Z$. 

For light quarks nonperturbative
corrections to $Z\gamma\gamma*$ are given by powers of $\Lambda_{\rm
QCD}^2/Q^2$ while perturbative ones are absent as we discussed
above. Thus, in the range of $Q$ of order $m_Z$ the one-loop
perturbative approach applies and suppression of the dependence on
$m_Z$ due to the cancellation of the $\log m_Z$ terms for $a_\mu^{\rm EW}$
between leptons and quarks in the first two generations is guaranteed.

The actual interplay of nonperturbative effects for light quark
contributions to $a^{\rm EW}_\mu$ represents an interesting picture
very different for the longitudinal $w_L$ and transversal $w_T$ parts.
For the first generation, in the chiral limit ($m_{u,d}=0$)
nonperturbative effects are absent in $w_L$ and the $1/Q^2$ one-loop
behavior in hadronic terms matches the massless pion pole. This is the
't Hooft matching condition, \cite{tHooft} as was pointed out in
\cite{Peris:1995bb,Knecht:2002hr}.  However, nonperturbative effects
are crucial for $w_T$, where they are responsible for a transformation
of the $1/Q^2$ singularity at small $Q$ into $\rho$ and $a_1$ meson
poles.

The situation is similar but somewhat more cumbersome for the $s$
quark in the second generation due to the U(1) anomaly (the
$\eta^\prime$ meson should be included together with $\eta$
meson). Also chiral breaking effects due to $m_s$ are more important.

Below we present a detailed discussion of perturbative and
nonperturbative effects for different generations.

\subsection{Third generation effect for \boldmath{$a^{\rm EW}_\mu$}}
\label{3gen} 

As we discussed above the one-loop expressions in (\ref{wlt})
work very well for the third generation where both perturbative and
nonperturbative corrections due to strong interactions are
small. Substituting $w_{L,T}$ from Eq.\,(\ref{wlt}) into (\ref{amu}) 
we get for the sum of $\tau $, $b$ and $t$ contributions to $a^{\rm EW}_\mu$
the following result \cite{Peris:1995bb,CKM95,Knecht:2002hr} 
\begin{eqnarray}
\hspace{-7mm} \Delta a^{\rm EW}_\mu[\tau, b,t]= -\frac{\alpha}
{\pi}\,\frac{G_\mu\,m_\mu 
^2}{8\pi ^2\sqrt{2}}
\left[\frac{8}{3}\ln\frac{m_t^2}{m_Z^2}-
\frac{2}{9}\,\frac{m_Z^2}{m_t^2}\left
( \ln\frac{m_t^2}{m_Z^2}+\frac{5}{3}\right)
+4\ln\frac{m_Z^2}{m_b^2}+3\ln\frac{m_b^2}{m_\tau 
^2}-\frac{8}{3}\right],
\end{eqnarray}
where we neglected small corrections  of order
$m_\mu^2/m_{\tau,b,t,Z}^2$,
$m_{\tau,b}^2/m_Z^2$, and $m_Z^4/m_t^4\,$. Numerically,
\begin{equation}
\Delta a^{\rm EW}_\mu[\tau, b,t]= -\frac{\alpha}{\pi}\,\frac{G_\mu\,m_\mu 
^2}{8\pi ^2\sqrt{2}}\cdot 30.3 =-8.21\cdot 10^{-11}\,,
\end{equation}
which is properly included in the
results of Eqs.~(\ref{eq17}) and  (\ref{eq18}).

Following the discussion in Sec.~\ref{hadroncor}, we estimate a
perturbative uncertainty by adding terms of order of
$\alpha_s(m_q)/\pi$ to $\log m_q$. It gives for the uncertainty
\begin{eqnarray}
-\frac{\alpha}{\pi}\,\frac{G_\mu\,m_\mu 
^2}{8\pi ^2\sqrt{2}}
\left\{\frac{16}{3}\,C_t \,\frac{ \alpha_s(m_t)}{\pi} 
-2\, C_b\, \frac{ \alpha_s(m_b)}{\pi} 
\right\},
\end{eqnarray}
where $C_t$ and $C_b$ are numbers of order $\pm 1$. Using 
$\alpha_s(m_Z)=0.11$, we come to an estimate
\begin{eqnarray}
\pm \frac{\alpha}{\pi}\,\frac{G_\mu\,m_\mu 
^2}{8\pi ^2\sqrt{2}}\cdot 0.3\approx \pm 0.1\cdot 10^{-11}\,.
\end{eqnarray}
\subsection{First and second generations: logarithmic estimates}
\label{sec:logest}

In the case of the first generation, i.e. $u$ and $d$ quark loops
together with the electron loop, the characteristic hadronic scales
are provided by the $\rho $ meson mass, $m_\rho =770\, {\rm MeV}$,
(for the vector current) and by the $a_1$ meson mass, $m_{a_1}=1260\,
{\rm MeV} $, (for the transversal part of the axial current).
Therefore, for $Q$ below $m_\rho$, only the electron loop contributes
to $w_T$.

On the other hand, the longitudinal part of the axial current is
dominated by the $\pi$ meson whose mass is small. This dominance means
that the one-loop expression (\ref{wlt}) for $w_L [e,u,d]$ (but not
for $w_T [e,u,d]$) works all the way down up to $Q\sim m_\pi $.  Thus,
the contribution of $w_L[e,u,d]$ in $a^{\rm EW}_\mu$ is strongly suppressed.

Considering $\ln(m_\rho ^2/m_\mu ^2)$ as a large parameter we see that with
logarithmic accuracy the first generation gives for $a^{\rm EW}_\mu$
\begin{equation}
\Delta a^{\rm EW}_\mu [e,u,d]=\frac{\alpha}{\pi }\,\frac{G_\mu\,m_\mu 
^2}{8\pi^2\sqrt{2}}
\int_{m_\mu^2}^{m_\rho ^2} {\rm d}Q^2
\,w_T^e=-\frac{\alpha}{\pi }\,\frac{G_\mu\,m_\mu ^2}{8\pi
^2\sqrt{2}}\ln\frac{m_\rho ^2}{m_\mu^2}=-1.08\cdot  10^{-11}\,,
\end{equation}
where we do not differentiate between $m_\rho$ and $m_{a_1}$.

The case of the second generation, $\mu, c, s$, is more involved.  The
cancellation of fermion loops takes place at $Q^2 > 4m_c^2$; so, we take
$m_{J/\psi }=3097\, {\rm MeV}$ as an upper limit of integration, and
$m_\phi =1019 \, {\rm MeV}$ plays the role of $m_\rho$ for the strange
quark.  In the interval between $m_\phi$ and $m_{J/\psi }$ the muon
and $s$ quark loops should be included in the integration over $Q$. In the
interval between $m_\eta=547\, {\rm MeV}$ and $m_\phi $ only the muon
loop contributes to $w_T$ but we need to account for the
pseudogoldstone nature of the $\eta$ meson. In contrast to the
first generation where the longitudinal part of the axial current had
the same quantum numbers as the $\pi$ meson, we need to re-express the
axial current of the strange quark as a combination of the SU(3)
singlet and octet,
\begin{equation}
j^{5}_\nu [s]= -\bar s\gamma _\nu \gamma _5 s = -\frac{1}{3}
(\bar u\gamma_\nu
\gamma _5 u+\bar d\gamma _\nu \gamma_5 d +
\bar s\gamma_\nu\gamma _5 s) + \frac{1}{3}(\bar u\gamma _\nu
\gamma _5 u+\bar d\gamma _\nu \gamma_5 d -2\bar s\gamma_\nu\gamma
_5 s)\,.
\end{equation}
The singlet part associated with $\eta'$ does not contribute to $w_L$ below
$m_{\eta'}$ (which is of the same order as $m_\phi$) but the octet
part  does, so $u$, $d$ and $s$ loops should
be taken with octet weights. In the last range, between $m_\mu $ and $m_\eta 
$
only the muon loop should be counted. Thus, we arrive at
\begin{eqnarray}
\Delta a^{\rm EW}_\mu [\mu,s,c]&=&\frac{\alpha}{\pi }\,\frac{G_\mu\,m_\mu 
^2}{8\pi^2\sqrt{2}}
\left\{\int_{m_\phi^2}^{m_{J/\psi }^2} {\rm d}Q^2
\,(w_L[\mu ,s]+w_T[\mu ,s])
\right.\nonumber\\[1mm]
&&
\hspace{-2.6cm}
\left.
+
\int^{m_\phi^2}_{m_\eta^2} {\rm d}Q^2
\left(\frac{2}{3}w_L[s]+{w_L[u]\over 3}-{w_L[d]\over 3}
+w_L[\mu]+
w_T[\mu] \right)
+
\int^{m_\eta^2}_{m_\mu^2} {\rm d}Q^2
\left(w_L[\mu] +w_T[\mu] \right)\right\}\nonumber\\[1mm]
&=&-\frac{\alpha}{\pi }\,\frac{G_\mu\,m_\mu ^2}{8\pi ^2\sqrt{2}}
\left(4\ln\frac{m_{J/\psi}^2}{m_\phi^2}
 +\frac{5}{3}\ln\frac{m_\phi^2}{m_\eta^2}
 +3\ln\frac{m_\eta^2}{m_\mu^2}
\right)=-5.64\cdot 10^{-11}\,.
\label{eq27}
\end{eqnarray}

To go beyond the logarithmic approximation, we must account for higher
powers of $1/Q^2$ for the light quarks.  The Operator Product
Expansion (OPE), which we next discuss, provides a systematic approach
for computing them. It leads to refined estimates of hadronic effects.

\subsection{OPE considerations}
\label{sec:OPE}

At large Euclidean $q^2$ one can use the OPE for the T-product of
electromagnetic and axial currents,
\begin{equation}
\label{ope}
\hat T_{\mu \nu }=i\int {\rm d}^4 x\, {\rm e}^{iqx}\,T\{j_\mu (x)\, j^5_\nu 
(0)\}=
\sum_i c^i_{\mu \nu \alpha_1\ldots\alpha_i}(q)\,
{\cal O}_i^{\alpha_1\ldots\alpha_i}\,,
\end{equation}
where the local operators ${\cal O}_i^{\alpha_1\ldots\alpha_i}$ are
constructed from the light fields.  A normalization point $\mu$, which
the operators and coefficients implicitly depend on, separates short
distances (accounted for in the coefficients $c^i$) and large
distances (in matrix elements of ${\cal O}_i$).  The field is light if
its mass is less than $\mu$. In the problem under consideration this
includes the electromagnetic field of the soft photon which can enter
local operators in the form of its gauge invariant field strength
$F_{\alpha\beta }=\partial _\alpha A_\beta -\partial _\beta A_\alpha
$. It also includes gluonic fields as well as lepton and quark fields
(with masses less than $\mu $).

The amplitude $T_{\mu \nu }$ is given by the matrix element of $\hat
T_{\mu \nu }$ between the photon and vacuum states,
\begin{equation}
\label{opemat}
T_{\mu \nu }=\langle 0|\,\hat T_{\mu\nu }\,|\gamma (k)\rangle=
\sum_i c^i_{\mu \nu \alpha_1\ldots\alpha_i}(q)\,
\langle 0|\,{\cal O}_i^{\alpha_1\ldots\alpha_i}\,|\gamma (k)\rangle\,.
\end{equation}
Since our approximation retains only terms linear in $k$, the matrix
elements are linear in $f_{\mu \nu }\!=\!k_\mu e_\nu\! -\!k_\nu e_\mu
$. This means that only ${\cal O}_i$ transforming under Lorentz
rotations as $(1,0)$ and $(0,1)$ contribute.  In other words, the
contributing operators should have a pair of antisymmetric indices,
${\cal O}_i^{\alpha\beta}=-{\cal O}_i^{\beta\alpha}$.
The amplitude $T_{\mu \nu }$ is a pseudotensor so it is convenient to
choose ${\cal O}_i^{\alpha\beta}$ to be a pseudotensor as well,
which is always possible using a convolution with $\epsilon _{\mu \nu
\gamma \delta }$.  Moreover, the $C$-parity of ${\cal O}^{\alpha\beta}$ 
should be $-1\,$.  Retaining only the
contributing operators we can write the OPE expansion as
\begin{equation}
\hat T_{\mu \nu }\!=\!\sum_i\left\{
c^i_T(q^2) \!\left(-q^2 {\cal O}^i_{\mu \nu}\!+\!q_\mu q^\sigma
{\cal O}^i_{\sigma  \nu }\! -\! q_\nu q^\sigma {\cal O}^i_{\sigma  \mu 
}\right)
\!+\!c^i_L(q^2)\, q_\nu q^\sigma {\cal O}^i_{\sigma  \mu}\right\}.
\end{equation}
Parametrizing
the matrix elements as
\begin{equation}
\langle 0|\,{\cal O}_i^{\alpha\beta}\,|\gamma 
(k)\rangle=-\frac{ie}{4\pi^2}\,\kappa
_i\,\tilde f^{\alpha\beta}\,,
\end{equation}
where the constants $\kappa _i$ depend on the normalization point $\mu $,
we get for the invariant functions $w_{T,L}$
\begin{equation}
w_{T,L}(q^2)=\sum c^i_{T,L}(q^2)\, \kappa_i\,.
\label{kappa}
\end{equation}

\subsubsection{The leading $d=2$ operator}

Operators are ordered by their canonical dimensions $d_i$ which define
the inverse power of $q$ in the OPE coefficients.  The leading
operators have minimal dimensions. In our problem the leading operator
dimension is $d=2$,
\begin{equation}
{\cal O}_F^{\alpha\beta}=\frac{e}{4\pi^2}\,\tilde F^{\alpha\beta}
=\frac{e}{4\pi^2}\,
\epsilon^{\alpha\beta \rho \delta }\partial
_\rho A_\delta\,,
\label{opF}
\end{equation}
where the operator $\tilde F^{\alpha\beta}$ is the dual of the
electromagnetic field strength, and the numerical factor is chosen in such
a way that $\kappa_{F}=1$.  Its OPE coefficients appear at one
loop. Note, that this operator corresponds to the unit operator in the
product of the three currents in (\ref{Tmunu1}). 

In considering the third generation, the normalization point $\mu$ can
be chosen well below all the masses $m_{\tau ,b,t}$; such that the
coefficients $c^F_{T,L}$ are given by a one-loop calculation,
\begin{equation}
c^F_{T,L}[\tau, b, t]=w_{T,L}[\tau, b, t]
\end{equation}
where $w_{L,T}[\tau ,b,t]$ are given by Eq.\,(\ref{wlt}) with summation only
over $f=\tau ,b,t$. Indeed, taking matrix element of $ \hat T_{\mu \nu}$
we are back to the one-loop expression for the amplitude $T_{\mu\nu}\,$.
Perturbative corrections are governed by $\alpha_s(m_b)$, nonperturbative 
ones are of order $\sim (\Lambda_{\rm QCD}/m_b)^4$ (due to the operator $F G 
G$ of dimension 6, see the discussion below).

Such a choice of the normalization point is not possible for the light
quarks $u,\,d,\, s$\,: to apply  the perturbative analysis to OPE
coefficients we should choose $\mu\gg \Lambda_{\rm QCD}\,$, i.e.\ 
the normalization point $\mu$ is certainly much
higher than the   light quark  masses. On the other hand, for
$Q\gg \Lambda_{\rm QCD}$ we can choose $\mu \ll Q\,$. For this range
$\Lambda_{\rm QCD}\ll \mu \ll Q$ we can use perturbation theory to
calculate the OPE coefficients. In particular, for the leading
operator (\ref{opF}) the OPE coefficients $c^F_{T,L}[f]$ in the chiral
limit $m_f=0$ are given by the one-loop expressions  for
$w_{T,L}[m_f=0]$  (see Eq.\,(\ref{msquare}) at $m_f=0$). An
interesting point here refers to dependence on $\mu$
for $c^F_{T,L}[m_f\!=\!0]$. This dependence is absent  not only at the
simple level of logarithmic corrections but also 
at the level of power  corrections in $\mu^2/Q^2$.

To demonstrate this let us consider the one-loop calculation of $\hat
T_{\mu\nu}$ in the background field method. Using the propagator
(\ref{prop}) at $m=0$ and doing the spinorial trace we come to the
following  expression 
\begin{equation}
  \label{oneback}
  \int \frac{{\rm d}^4 p}{p^4 (p+q)^2}\left[(p+q)_\mu\, p^\rho 
   \tilde F_{\rho\nu} + (p+q)_\nu \, p^\rho  \tilde F_{\rho\mu}\right],
\end{equation}
where we have omitted an unimportant overall factor. The integral over
Euclidean virtual momentum $p$ is well defined both in infrared and in
ultraviolet domains (the logarithmic divergence at large $p$ drops
out because of antisymmetry of $ F_{\mu\nu}$). It means that the
dominant contribution comes from the range of $p \sim q$, since there is no
other scale. A simple integration results in 
\begin{equation}
  \label{oneback1}
  \frac{1}{q^2}\left[q_\mu\, q^\rho 
   \tilde F_{\rho\nu} + q_\nu \, q^\rho  \tilde F_{\rho\mu}\right].
\end{equation} 
To this one has to add the loop with the massive
Pauli-Vilars regulators which is simple to calculate using the propagator 
(\ref{prop}). This contribution adds a polynomial term 
$-\tilde F_{\mu\nu}$  to the expression (\ref{oneback1}) what restores
the transversality for the electromagnetic current $j_\mu $ and leads
to the results (\ref{msquare}) for $w_{T,L}$ at $m_f=0$. 

The calculation above proves that up to  power accuracy the OPE
coefficients \mbox{$c^F_{T,L}[m_f\!=\!0]=w_{T,L}[m_f\!=\!0]\,$.} 
The portion of the integral (\ref{oneback}) which comes from 
the range $|p|<\mu$ constitutes a correction of order $\mu^2/Q^2$. 
However, even this correction is absent if the symmetry
features  of the theory are preserved. Indeed, the polynomial part 
$-\tilde F_{\mu\nu}$ which came from the regulator loop is due to 
arbitrary short distances so it does not depend on $\mu$ at all. 
The conservation of the electromagnetic current fixes the coefficient
between the polynomial regulator part and the dispersive part
(\ref{oneback1}).   
Separating the part of the integral (\ref{oneback}) with $|p|<\mu$ we
break the conservation. 
Moreover, due to chiral symmetry there are no other kinds of corrections, 
perturbative or nonperturbative,  to the one-loop result
for the longitudinal coefficient $c^F_L[m_f\!=\!0]\,$. That property
is completely valid
when we deal with the flavor nonsinglet axial current, like $\bar u \gamma_\nu
\gamma_5 u - \bar d \gamma_\nu\gamma_5 d$ in the first generation,
which has no gluonic U(1) anomaly. For the invariant function
$w_L[m_f\!=\!0]$ that feature follows from the Adler-Bardeen theorem
\cite{Adler:er} and 't~Hooft matching condition \cite{tHooft}. 
In terms of the OPE coefficients it translates in the equality of 
$c^F_L[m_f\!=\!0]$ and $w_L[m_f\!=\!0]$ because all other OPE
coefficients vanish at $m_f=0$ for the longitudinal part. 

For the transversal coefficient $c^F_T[m_f\!=\!0]$ the situation is
more subtle. As shown in \cite{WT} the symmetry of the dispersive part
(\ref{oneback1}) under the $\mu$, $\nu$ permutation leads to the
perturbatively exact relation $2c^F_T[m_f\!=\!0]=c^F_L[m_f\!=\!0]\,$. 
Nonperturbatively this relation is broken at the level of
$\Lambda_{\rm QCD}^4/Q^4$ terms as we will see from the OPE analysis.

We can also account for corrections to $c^F_{T,L}$ due to fermion
masses which break the chiral symmetry. They can be read off
Eq.\,(\ref{msquare}) for $w_{T,L}\,$. The corrections are of the
second order in $m_f$ but logarithmically sensitive to the
normalization  point $\mu$ which replaces $m_f$ under the log in
Eq.\,(\ref{msquare}) when we translate to $c^F_{T,L}\,$,
\begin{equation}
\label{cF}
c^F_{L}[f]=2c^F_{T}[f]=\frac{4I^3_f \,N_f\,Q_f^2}{Q^2}
\left[1-\frac{2\,m^2_f}{Q^2}
\ln\frac{Q^2}{\mu^2}
+{\cal O}\left(\frac{m^4_f}{Q^4}\right)\right].
\end{equation}
Summation over $f$, say for the first generation $e,\,u,\,d$,  leads
to the lepton-quark cancellation of the leading $1/Q^2$ terms and we must 
consider operators of higher dimensions.

\subsubsection{Operators of higher dimension}
 
The next operators, by dimension, are those of $d=3$
\begin{equation}
{\cal O}^{\alpha\beta}_{f}=-i\,\bar f 
\sigma^{\alpha\beta}\gamma _5 f 
\equiv \frac{1}{2}\,\epsilon^{\alpha\beta\gamma\delta}
\bar q_f\,\sigma_{\gamma\delta}\, q^f\,.
\label{fop}
\end{equation}
Chirality arguments
show that their OPE coefficients $c^f$ contain mass $m_f$ as a factor, so
by dimension, $c^f\propto m_f/Q^4$. To calculate these coefficients it is
sufficient to consider tree diagrams of the Compton scattering type,
\begin{equation}
c^f_L=2c^f_T=\frac{8I^3_f \,Q_f m_f}{Q^4}\,.
\end{equation}
Taking matrix elements between the soft photon and vacuum
states we produce 
the following terms in the invariant functions $w_{T,L}(q^2)$:
\begin{eqnarray}
\label{d3}
\Delta^{(d=3)} w_L=2\,\Delta^{(d=3)}w_T=\frac{8}{Q^4}\sum_f I^3_{f}\,Q_f m_f
\kappa _f \,.
\label{d3te}
\end{eqnarray}

If we neglect effects of strong interactions, it is simple to
calculate $\kappa _f$ in one loop with logarithmic accuracy using,
e.g., the propagator (\ref{prop}) in the external field and      
the normalization   point $\mu$  as the UV regulator,
\begin{equation}
\kappa _f=-Q_f\,N_f\, m_f \ln\frac{\mu^2}{m_f^2}\,.
\end{equation}
Substituting this $\kappa_f$ in Eq.\,(\ref{d3}) we observe the full match with
the $1/Q^4$ term in Eq.\,(\ref{cF}), together the $d=2$ and $d=3$ 
operators reproduce the one-loop $w_{L,T}$ in Eq.\,(\ref{msquare}).

Note that this match is for the terms of second order in mass.  In QCD
due to spontaneous breaking of chiral symmetry the matrix elements of
quark operators (\ref{fop}) are not vanishing at $m_q=0$, instead they
are proportional to the quark condensate $\langle \bar q
q\rangle_{_0}=-(240\,{\rm MeV})^3\,$.  The operators (\ref{fop}) played
an important role in the analysis by Ioffe and Smilga
of nucleon magnetic moments with QCD sum rules \cite{Ioffe:1984ju}. 
They determined by a sum rule fit the quantity 
\begin{eqnarray}
\label{magmom}
\chi=-\frac{\kappa_q}{4\pi^2\, Q_q\,\langle \bar q
  q\rangle_{_0}}=-\frac{1}{(350 \pm 50~\mbox{MeV})^2} 
\end{eqnarray}
dubbed as the quark condensate magnetic susceptibility.

Actually, the OPE analysis together with the pion dominance in the
longitudinal part leads to a relation for magnetic susceptibility 
similar to the Gell-Mann-Oakes-Renner (GMOR) relation for the pion
mass \cite{GMOR}. This relation derived in Ref.~\cite{WT} has the form
\begin{eqnarray}
  \label{magsuc}
  (m_u+m_d)\,\kappa_q=- m_\pi^2\,N_c\,Q_q
\end{eqnarray}
The GMOR relation $F_\pi^2 \,m_\pi^2= -(m_u+m_d)\langle \bar q
q\rangle_{_0}$ allows us to rewrite (\ref{magsuc}) as
\begin{eqnarray}
  \label{magsuc1}
\kappa_q=-4\pi^2\,Q_q \,\langle \bar q q\rangle_{_0}\, \chi\,,
\qquad \chi=-\frac{N_c}{4\pi^2 \,F_\pi^2}
=-\frac{1}{(335~\mbox{MeV})^2} \,.
\end{eqnarray}
Although this value of $\chi$ is in a good agreement with the QCD sum rule fit
(\ref{magmom}) its magnitude is about two times higher than results of other
approaches, see \cite{WT} for references and discussion. 

Let us go further by operator dimensions. Nothing new appears for
$d=4$: all operators of dimension 4 are
reducible to the $d=3$ operators
due to the following relation,
\begin{equation}
 \bar f (D_\mu \gamma _\nu -
D_\nu \gamma _\mu)\gamma _5 f=-m_f \bar f \sigma _{\mu \nu}\gamma _5 f\,.
\end{equation}

\begin{figure}[ht]
\hspace*{0mm}
\begin{minipage}{16.cm}
\begin{tabular}{c@{\hspace{16mm}}c}
\psfig{figure=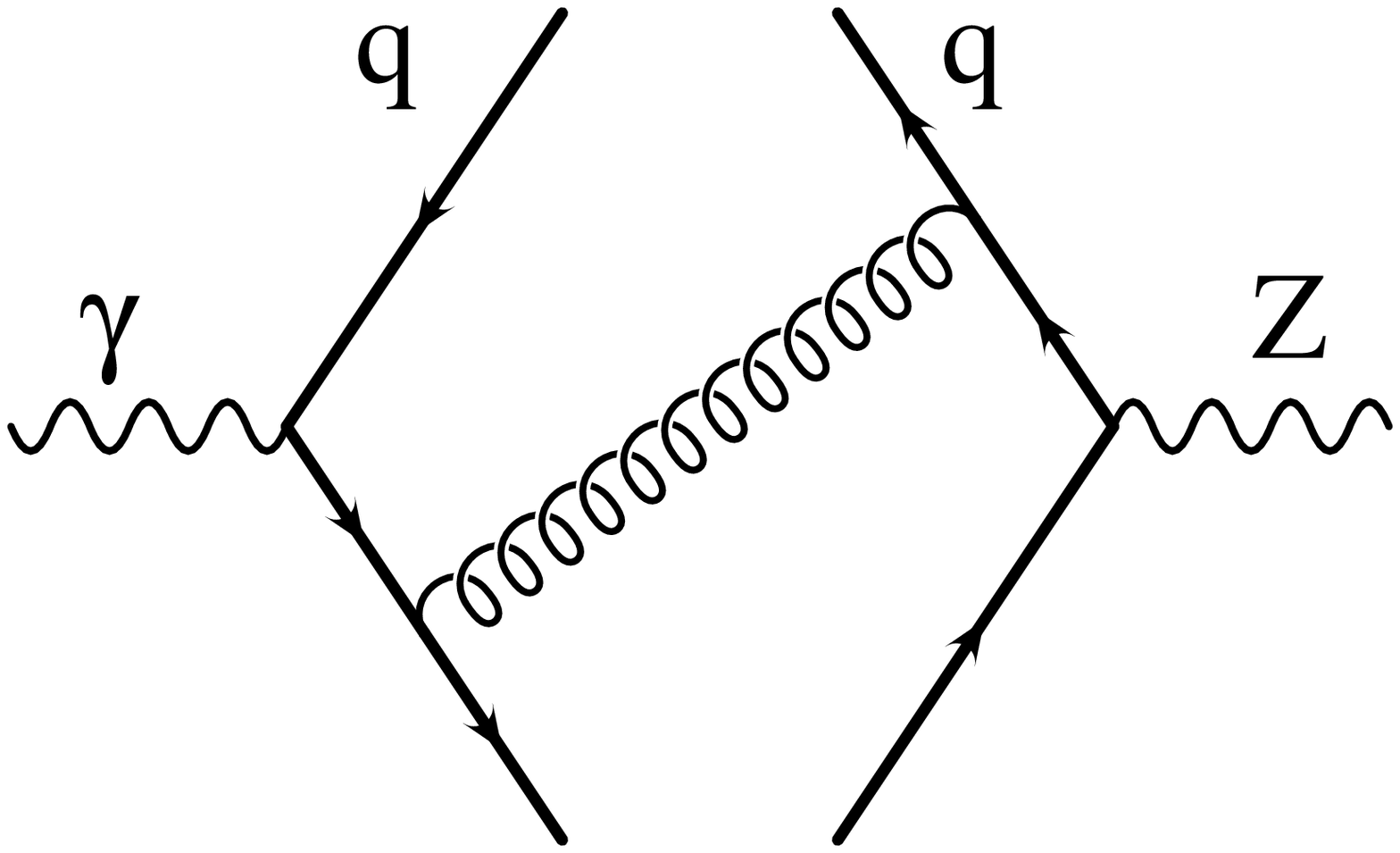,width=48mm}
&
\psfig{figure=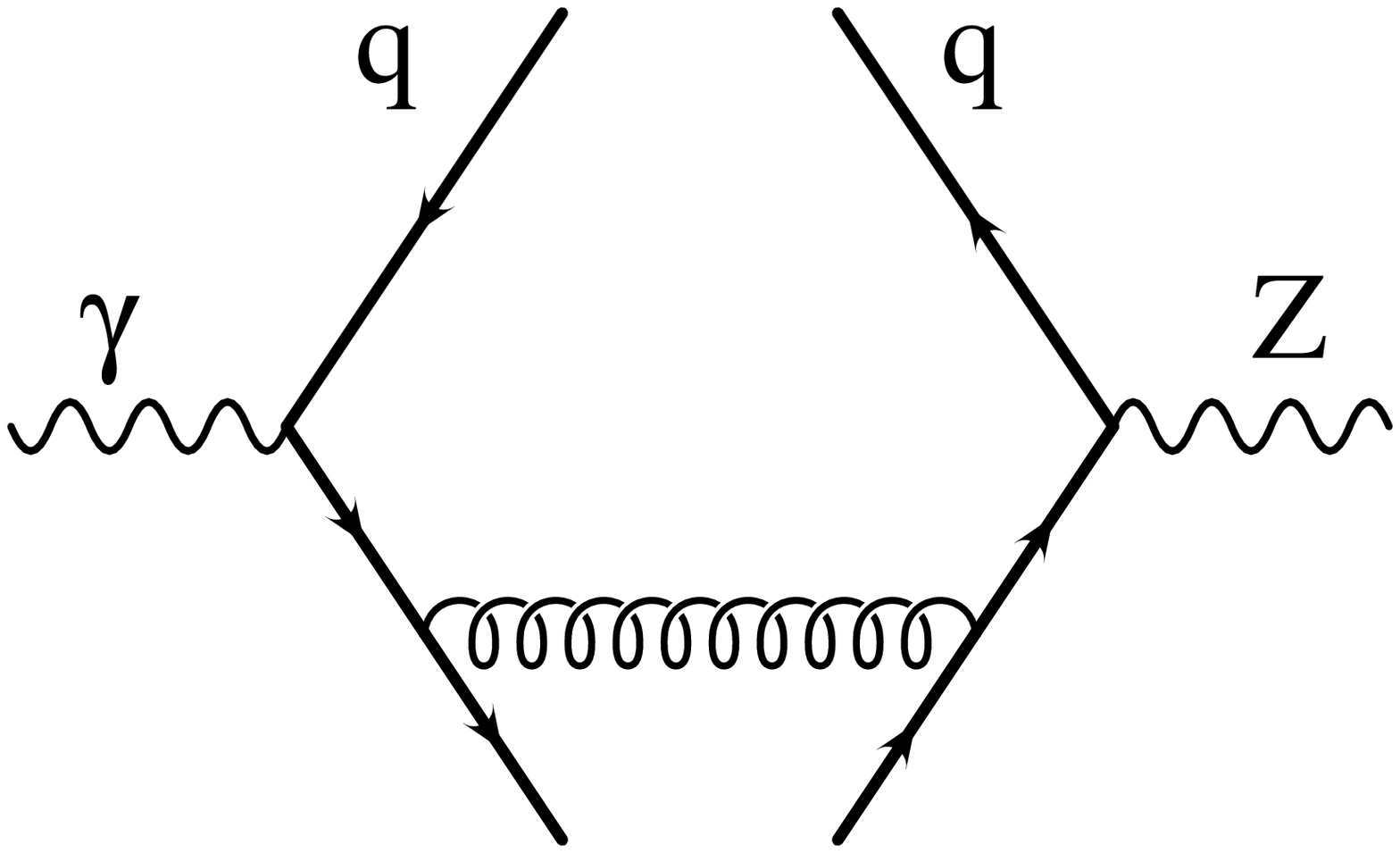,width=48mm}
\end{tabular}
\end{minipage}
\caption{\sf \small Diagrams for four-fermion operator ${\cal O}_6\,$.}
\label{fig:fourF}
\end{figure}

For $d=5$ we have operators
$\bar f f \tilde F^{\alpha\beta}$ and $\bar f\gamma _5 f \tilde
F^{\alpha\beta}$ (with  factors $m_f$ again) and at $d=6$ there are
many operators of the type $(\bar f \sigma ^{\alpha\beta} 
\gamma _5 f) \,(\bar f f)$, $\tilde F^{\alpha\beta} {\rm
Tr}\, G_{\mu \nu }G^{\mu \nu }$ and so on. These $d=5,6$ operators
produce $1/Q^6$ terms in $T_{\mu \nu }$. A particular example is the
following four-fermion operator which appears due to diagrams in
Fig.~\ref{fig:fourF},
\begin{equation}
\label{omunu}
{\cal O}_6^{\alpha\beta}= \bar q \gamma^{\alpha}\gamma_5\, Q\,
t^a q \; \bar q \gamma^{\beta}\,I^3\,t^a q -(\alpha
\leftrightarrow \beta) ,
\end{equation}
where the quark field $q$ has color and flavor indices, the matrices 
of color generators 
$t^a$, $a=1,\ldots,8$ act on color indices, and the weak isospin $I^3$
and electric charge $Q$ are the diagonal matrices in the flavor space.
This operator  enters in the 
OPE (\ref{opemat}) with the following coefficients
\begin{equation}
c^6_T= -\frac{16\pi \alpha_s(Q)}{Q^6}\,, \qquad c^6_L=0\,.
\end{equation}

Note that our consideration of four-fermion operators is similar to 
Ref.~\cite{Knecht:2002hr}.  
Note also that the operator $O_6$ contributes only to the transversal
function $w_T$ --- consistent with absence of nonperturbative
corrections to $w_L$ in the chiral limit. Moreover, in the chiral
limit,  the $d=6$ operators are next-to-leading after the leading $d=2$
operator, showing that parametrically the leading nonperturbative
corrections are of order $\Lambda_{\rm QCD}^4/Q^4\,$.

The matrix element of (\ref{omunu}) between the vacuum and the photon
states can be found assuming factorization in terms of the quark
condensate $\langle \bar q q\rangle_{_0}$ and the magnetic
susceptibility $\kappa_q$ given in Eq.\,(\ref{magsuc1}). It
results in the following piece in $w_T$ for the light quarks in the
first ($u,d$) and second ($s$) generations,
\begin{eqnarray}
&&\Delta^{(d=6)} w_T[u,d]=-3\Delta^{(d=6)} w_T[s]=
-\frac{32 \pi\alpha_s}{9\,Q^6} \,\frac{\langle \bar q q\rangle_{_0}^2}
{F_\pi^2}=-\alpha_s\, \frac{(0.71\; {\rm GeV})^4}{Q^6}\,.
\end{eqnarray}
We can use this as an estimate for the $1/Q^6$ terms in $w_T$
neglecting the $F G G$ operators which enter with smaller
coefficients (they appear in one loop while ${\cal O}_{6}$ is due
to tree level diagrams). We also neglect the anomalous dimension of
${\cal O}_6$. However, we have in mind that this anomalous dimension
is rather large and positive and considerably compensates the running
of $\alpha_s$ which therefore can be taken close to 1 for estimates.

Summarizing the consequences of OPE for the $u$,  $d$ and $s$ quark loops in
the chiral limit we get 
\begin{eqnarray}
\label{wud}
&& w_L[{u,d}]_{m_{u,d}=0}=-3w_L[s]_{m_{s}=0}=\frac{2}{Q^2}\,,
 \nonumber\\[1mm]
&&w_T[{u,d}]_{m_{u,d}=0}=-3w_T[s]_{m_{s}=0}=
\frac{1}{Q^2} -\frac{(0.71\; {\rm
GeV})^4}{Q^6}+{\cal O}\left(\frac{1}{Q^8}\right).
\end{eqnarray}
The longitudinal part given by the leading $d=2$ operator $\tilde
F_{\mu\nu}$ has neither perturbative $\alpha_s$ corrections nor
nonperturbative ones, and the pole $1/Q^2$ matches the massless
pion. So the cancellation of $w_L^{u,d}$ and $w_L^{e}$ in the first
generation is exact in the chiral limit $m_{u,d}=m_e=0$.

As we discussed above the leading operator contribution to the
transversal part has no perturbative corrections either  \cite{WT}; however,
nonperturbative corrections are present. Their signal in the chiral
limit is very clean: the lowest masses in the vector and axial vector
channels are nonvanishing in contrast with the pion in the
longitudinal part. In Sec.\,\ref{sub:first} below we present a resonance model
for $w_T[u,d]$ consistent with the OPE constraints.  Thus, there is 
no complete cancellation in the sum
of $w_T^{u,d}$ and $w_T^{e}$ although this sum decreases as $1/Q^6$ at
large $Q$.

\subsubsection{Comparison with the OPE analysis 
in Ref.\,\cite{Knecht:2002hr}}

It is convenient at this point to discuss a comparison of our approach
with the analysis of the $u$, $d$ quark loops of
Ref.~\cite{Knecht:2002hr}. In essence,
it is claimed there  that the leading large $Q$ behavior of the 
transversal
part $w_T[{u,d}]$ (and  $w_T[{s}]$ as well) is
\begin{equation}
w_T[{u,d}]\propto \frac{1}{Q^6}
\end{equation}
in the chiral limit in contrast with the $1/Q^2$ perturbative
behavior.
As a result, the $1/Q^2$ part of $w_T[{u,d}]$ is absent while 
the $1/Q^2$ part of $w_T[{e}]$ is present, so the quark-lepton
cancellation is destroyed and, consequently, spurious 
$\log m_Z$ terms appear in $a^{\rm EW}_\mu$. 

To pinpoint the origin for such a dramatic difference between our approaches 
let us notice first that in \cite{Knecht:2002hr} the
authors consider $T_{\mu\gamma\nu}$, the vacuum average of the product
of three currents defined in Eq.\,(\ref{Tmunu1}), as a primary
object. In this approach they do not have the electromagnetic field
entering into local operators and our leading operator
$\tilde F_{\alpha\beta}$ does not appear.
However, they have to consider the OPE for the product of three currents
instead of two. For  $T_{\mu\gamma\nu}$ one can
derive the following expansion: 
\begin{equation}
  \label{munuga}
  T_{\mu\gamma\nu}=\langle\, 0\, |\,\sum_i 
c^i_{\mu\gamma\nu\alpha_1\ldots\alpha_i}(q,k)\,
{\cal O}_i^{\alpha_1\ldots\alpha_i}(0)+i\int {\rm d}^4 y\,{\rm e}^{-iky}\,
T\{\hat T_{\mu\nu}(0)\, j_\gamma(y)\}\, |\,0\,\rangle\,. 
\end{equation}
The first part, which contains the local operators, accounts for emission
of the soft photon from short distances.  The second bilocal part, which
contains the operator $\hat T_{\mu\nu}$ defined in Eq.\,(\ref{ope}) and
the soft momentum current $j_\gamma$, accounts for the soft photon
emission from large distances. Our relation (\ref {opemat})
conveniently includes  both parts in the local form: the first one is 
  due to operators ${\cal O}_i$ containing the electromagnetic field
strength $F_{\alpha\beta }$ explicitly and the second is  due to the
operators without $F_{\alpha\beta }\,$. Moreover, our leading operator
 $\tilde F_{\alpha\beta}$ corresponds to the unit operator in the
 first, local, part on the r.h.s.\ of Eq.\,(\ref{munuga}). 

Once we have established this correspondence it is simple to see what
is missing in the analysis of \cite{Knecht:2002hr}: they did not
account for the local part in Eq.\,(\ref{munuga}). It is the unit
operator in this part ($\tilde F_{\alpha\beta}$ in our formalism) which
gives the leading $1/Q^2$ contribution and its coefficient follows from
the perturbative triangle. This corresponds to soft photon emission
from short distances as we discussed above.

Note, that the unit operator in the local part of Eq.\,(\ref{munuga})
is leading for both, longitudinal and transversal structures in the
amplitude $T_{\mu\gamma\nu}$, so its omission  should
result in an error for the longitudinal contribution as well. 
This did not happen in \cite{Knecht:2002hr} because they did not
apply  the OPE analysis to the longitudinal part of the amplitude,
instead fixing it by the anomaly. A comparison with the expansion
(\ref{munuga}) is particularly simple for the longitudinal part at
$m_f\!=\!0\,$:
the second bilocal term containing $\hat T_{\mu\nu}$ vanishes and the
only surviving operator is the unit operator in the first term.

All the OPE subtleties discussed above should not screen a conceptually
very simple situation: the short distance behavior given by free
quark loops  should not be changed in QCD by large distance effects.

\subsection{First generation}
\label{sub:first}

We are now well prepared to calculate the contribution of the first
generation 
to $a^{\rm EW}_\mu$ with an accurate account of hadronic
effects. The invariant functions  $w_{T,L}$ for the first generation is
the sum of $w_{T,L}[e]$, $w_{T,L}[u]$ and
$w_{T,L}[d]\,$. For the electron
\begin{equation}
w_{L}[e]=2\,w_{T}[e]=-\frac{2}{Q^2}\,,
\end{equation}
where we neglected the electron mass.  Expansions at large $Q^2$ in
the chiral limit for $w_{T,L}[u,d]$ are given in Eq.\,(\ref{wud}).  Hadronic
effects modify $w_{T,L}[u,d]\,$. Modifications are minimal for the
longitudinal function $w_{L}[u,d]$: the position of the pole is
shifted to $m_\pi^2$ due to the explicit breaking of the chiral
symmetry by quark masses,\footnote{It is just this shift which
allows one to derive \cite{WT} the expression in (\ref{magsuc}) by
comparison of the $1/Q^4$ terms with the OPE.}
\begin{equation}
w_L[{u,d}]=\frac{2}{Q^2+m_\pi^2}\,.
\end{equation}
To find the contribution of $w_L[{u,d}]$ to $a^{\rm EW}_\mu$ one needs 
to use the more accurate Eq.\,(\ref{amug}), rather than 
Eq.\,(\ref{amu}), because
the integral is dominated by momenta $Q\sim m_\pi$ comparable with
$m_\mu$,
\begin{eqnarray}
\Delta a_\mu^L[e,u,d]&=&
- \frac{\alpha}{\pi}\,\frac{G_\mu 
m_\mu^2}{8\pi^2\sqrt{2}}
\left\{2\ln\frac{m_\pi^2}{m_\mu^2}+\frac{8}{3}
+\frac{4}{3}\int_0^1{\rm d}  
\alpha(1+\alpha)\ln A
\right.
\nonumber\\
&& \left. \qquad \qquad
+4\,\frac{m_\mu^2}{m_\pi^2}\left[\int_0^1{\rm d} 
\alpha(1-\alpha)^2\ln 
A-\frac{1}{3}\ln\frac{m_\mu^2}{m_\pi^2}+\frac{2}{9}\right]
\right\},
\end{eqnarray}
where $A=\alpha +(1-\alpha)^2(m_\mu^2/m_\pi^2)$. Numerically it gives
\begin{equation}
\label{alud}
\Delta a_\mu^L[e,u,d]=- \frac{\alpha}{\pi}\,\frac{G_\mu 
m_\mu^2}{8\pi^2\sqrt{2}}\cdot 3.53=
-0.96 \cdot 10^{-11}\,.
\end{equation}

The transversal function $w_{T}[{u,d}]$ can be modeled as a linear
combination of two pole terms: one is due to the $\rho(770)$ vector
meson, another due to the $a_1(1260)$ axial vector meson,
\begin{equation}
\label{wtud}
w_T[{u,d}]=\frac{1}{m_{a_1}^2-m_\rho^2}
\left[\frac{m_{a_1}^2-m_\pi^2}{Q^2+m_\rho^2}   
-\frac{m_\rho^2-m_\pi^2}{Q^2+m_{a_1}^2}\right]\,.
\end{equation}
The residues in this expression are fixed by two conditions at large
$Q$ which follow from the OPE expression (\ref{wud}) plus the $d=3$
terms (\ref{d3te}) breaking chiral symmetry.  The first condition is
on the coefficient of the leading $1/Q^2$ term, the second condition
is for the coefficient of $1/Q^4\,$. The term $1/Q^6$ in (\ref{wud})
allows for an extra test of the model.  The expression (\ref{wtud})
gives $-(0.96\,{\rm GeV})^4$ to be compared with $-(0.71\,{\rm
GeV})^4$ in the OPE based (\ref{wud}). Agreement is not extremely good
but the right sign and order of magnitude are encouraging. Since the
OPE $1/Q^6$ estimate is very approximate, we use (\ref{wtud}) for
numerical estimates.

For the integral over $Q$ defining the contribution of $w_T[{u,d}]$, we
can use the simpler expression (\ref{amu}) neglecting $m_\mu^2/m_\rho^2$
corrections,
\begin{equation}
\label{atud}
\Delta a_\mu^T[e,u,d]=- \frac{\alpha}{\pi}\,\frac{G_\mu 
m_\mu^2}{8\pi^2\sqrt{2}}
\left\{\ln\frac{m_\rho^2}{m_\mu^2}
-\frac{m_\rho^2}{m_{a_1}^2-m_\rho^2}\ln\frac{m_{a_1}^2}{m_\rho^2} 
+\frac{3}{2}
\right\},
\end{equation}
which gives numerically
\begin{equation}
\Delta a_\mu^T[e,u,d]=- \frac{\alpha}{\pi}\,\frac{G_\mu 
m_\mu^2}{8\pi^2\sqrt{2}}\cdot 4.88=
-1.32\cdot 10^{-11}\,.
\end{equation}

Overall, the first generation contributes to $a^{\rm EW}_\mu$
\begin{equation}
\label{atudn}
\Delta a^{\rm EW}_\mu [e,u,d]=- \frac{\alpha}{\pi}\,\frac{G_\mu 
m_\mu^2}{8\pi^2\sqrt{2}}\cdot 8.41=
-2.28\cdot 10^{-11}\,,
\end{equation}
which is to be compared with the constituent quark model result \cite{CKM95},
\begin{equation}
\label{freeq1}
\Delta a^{\rm EW}_\mu [e,u,d]_{\rm free~quarks}=- \frac{\alpha}{\pi}\,\frac{G_\mu 
m_\mu^2}{8\pi^2\sqrt{2}}\left[ \ln\frac{m_u^8}{m_\mu^6 m_d^2} 
+\frac{17}{2}\right]=
-4.0\cdot 10^{-11}\,.
\end{equation}
The refined result is about 1/2 of the constituent quark model value.
It represents our main phenomenological finding.  The primary reason
for the shift is a deeper extension into the infrared due to
quark-hadron duality for the longitudinal function $w_L[u,d]$. It
leads to a stronger quark-lepton cancellation for $w_L$ -- the effect
noted in Ref.~\cite{Knecht:2002hr}.

What is the accuracy of the result (\ref{atudn})? Most of the model
dependence is related to the description of the transversal function
$w_T[u,d]$.  For the longitudinal contribution, the analysis is rather
solid. To get an idea of the accuracy, we consider variations when the
$\rho$ and $a_1$ masses are in the intervals $500-1000$ MeV and
$900-2000$ MeV.  We found that deviations from the result
(\ref{atudn}) are within 10\%, i.e.\ of order of $\pm 0.2\cdot
10^{-11}\,$. This high level of stability is related to the fact that 
the main contribution to 
$\Delta a_\mu^T[e,u,d]$ in Eq.\,(\ref{atud}) comes from the unambiguous 
logarithmic term $\ln(m_\rho^2/m_\mu^2)$; it gives 
$-1.08\cdot 10^{-11}$ out of $-1.32\cdot 10^{-11}\,$.

So, in total, this analysis
increases $a_\mu^{\rm EW}$ by $2\times 10^{-11}$ relative to the free
quark calculation and significantly improves its
reliability.

\subsection{Second generation}

The second generation contains both light, $s\,$, and  heavy, $c\,$, quarks
which should be treated differently. For the light $s$ quark we use
the approach similar to the case of $u,d$ quarks. For the longitudinal
function $w_L[s]\,$, as it was explained above in 
Sec.~\ref{sec:logest}, one must include both singlet, $\eta'(960)\,$,
and octet, $\eta(550)$, pseudoscalar mesons,
\begin{equation}
\label{wls}
w_L[s]=-\frac{2}{3}\left[\frac{2}{Q^2+m_{\eta'}^2}-\frac{1}{Q^2+m_{\eta}^2}
\right].
\end{equation}
For the transversal function the model is
\begin{equation}
\label{wts}
w_T[s]=-\frac{1}{3}\,\frac{1}{m_{f_1}^2-m_\phi^2}
\left[\frac{m_{f_1}^2-m_\eta^2}{Q^2+m_{\phi}^2}-\frac{m_{\phi}^2
-m_\eta^2}{Q^2+m_{f_1}^2}
\right],
\end{equation}
where $\phi(1019)$ and $f_1(1426)$ are isoscalar vector and
axial vector mesons relevant to the $\bar s s$ channel.
Integrating $w_{L,T}[s]$ and adding the known expression for the  $c$ quark
and muon contribution we get for the second generation
\begin{eqnarray}
\label{secgen}
\Delta a^{\rm EW}_\mu[\mu,s,c]&=&- \frac{\alpha}{\pi}\,\frac{G_\mu 
m_\mu^2}{8\pi^2\sqrt{2}}\,\left\{ 
\frac{2}{3}\ln\frac{m_\phi^2}{m_{\eta'}^2}-
\frac{2}{3}\,\ln\frac{m_{\eta'}^2}{m_\eta^2}
+\frac{1}{3}\,\frac{m_\phi^2-m_\eta^2}{m_{f_1}^2-m_\phi^2}\,
\ln\frac{m_{f_1}^2}{m_\phi^2}\right.
\nonumber\\[1mm]
&&\hspace{2.4cm} \left.+\, 4\ln\frac{m_c^2}{m_\phi^2}
+3\ln\frac{m_\phi^2}{m_\mu^2}-\frac{8\pi^2}{9}+\frac{56}{9}\right\}.
\end{eqnarray}
Numerically it constitutes (at $m_c=1.5$ GeV)
\begin{eqnarray}
\label{secgenN}
\Delta a^{\rm EW}_\mu[\mu,s,c]=- \frac{\alpha}{\pi}\,\frac{G_\mu 
m_\mu^2}{8\pi^2\sqrt{2}}\cdot 17.1=-4.63\cdot 10^{-11}\,.
\end{eqnarray}
This numerical value practically coincides with the free quark
calculation \cite{CKM95},
\begin{equation}
\label{freeq2}
\Delta a^{\rm EW}_\mu [\mu,c,s]_{\rm free~quarks}=
- \frac{\alpha}{\pi}\,\frac{G_\mu 
m_\mu^2}{8\pi^2\sqrt{2}}\left[ \ln\frac{m_c^8}{m_\mu^6 m_s^2} 
+\frac{47}{6} -\frac{8\pi^2}{9}\right]=
-4.65\cdot 10^{-11}\,.
\end{equation}
Two reasons for such a good agreement. First is the smallness of 
the strange quark contribution -- its electric charge is smaller --  
so,  hadronic details are not so important. Second is that the effect
of cancellation between leptons and hadrons in the longitudinal
invariant function $w_L$ is much less pronounced that in the first
generation because of larger masses of $\eta$ and $\eta'\,$.

The result is more sensitive to the $c$ quark parameters. If we take
1.3 GeV for its mass instead of 1.5 we get $(-4.32)\cdot  10^{-11}$,
i.e.\ a change by $0.3\cdot  10^{-11}$. Another source of the
QCD corrections for the heavy quarks is perturbative gluon exchanges
in the quark triangles. This estimation is similar to to one we did in
Sec.~\ref{3gen}, we substitute $\log m_c$ by $\alpha_s(m_c)/\pi$,
\begin{eqnarray}
\label{pertgl}
 \frac{\alpha}{\pi}\,\frac{G_\mu 
m_\mu^2}{8\pi^2\sqrt{2}}\cdot 8\,\frac{\alpha_s(m_c)}{\pi}
\approx 0.2 \cdot 10^{-11}\,,
\end{eqnarray}
where we used $\alpha_s(m_c)\approx 0.3\,$.

We conclude that the uncertainty coming from the second generation is
small, about $\pm 0.3\times 10^{-11}$, and related mainly to charm
quark parameters.  Overall, the total hadronic loop uncertainties in
$a_\mu^{\rm EW}$ are well accounted for by an error of $\pm 1\times
10^{-11}$.

\section{Leading logarithms: renormalization group analysis}
\label{sec:logs}
It was pointed out in \cite{CKM96} that once the leading log
short-distance two-loop corrections to $a_\mu^{\rm EW}$ of order $
\ln (m_W/ m_\mu)$ are completely known, a renormalization group
(RG) analysis can provide all leading log terms of the form $\left[\alpha
\ln (m_W/m_\mu)\right]^n$, $n=2,3,\ldots\,$, coming from $n+1$ loop
effects.  Such an analysis was carried out in Ref.~\cite{Degrassi:1998es} for
the leading log three-loop contribution.  Since we have now clarified
the short-distance two-loop behavior of $a_\mu^{\rm EW}$, it is
appropriate for us to revisit the issue of higher orders and refine
the previous study.

An interesting subtle feature that enters this RG analysis is the
mixing of operators.  We are interested in the OPE coefficient of the
dimension 5 dipole operator $\bar \mu \sigma_{\alpha\beta} \mu \,
F^{\alpha\beta}$.  Leading logs contribute at the two-loop level due
to QED corrections to the dipole operator (its anomalous dimension) as
well as from two-loop mixing between the dipole operator and $d=6$,
four-fermion operators.  A careful treatment of their mixing is
important for the RG analysis. 

Before addressing the details of $a_\mu^{\rm EW}$, it is useful to
recall that a related QCD study was carried out about 25 years ago
\cite{Shifman:1977ge,Shifman:1978de} for the case of weak radiative
decays which involve flavor changing gluomagnetic and electromagnetic
dipole operators.  Indeed, there QCD effects are very large and a RG
analysis is essential.  Later, because of the phenomenological
importance of $b\to s\gamma$, interest in such transition dipole
operators increased, generating many studies and some controversies
involving subtle issues regarding renormalization scheme dependence,
$\gamma_5$ definition, operator set completeness etc.  A brief
discussion of those issues will provide guidance for our $a_\mu^{\rm
EW}$ three-loop analysis.

\begin{figure}[ht]
\hspace*{0mm}
\begin{minipage}{16.cm}
\begin{tabular}{c@{\hspace{16mm}}c@{\hspace{16mm}}c}
\psfig{figure=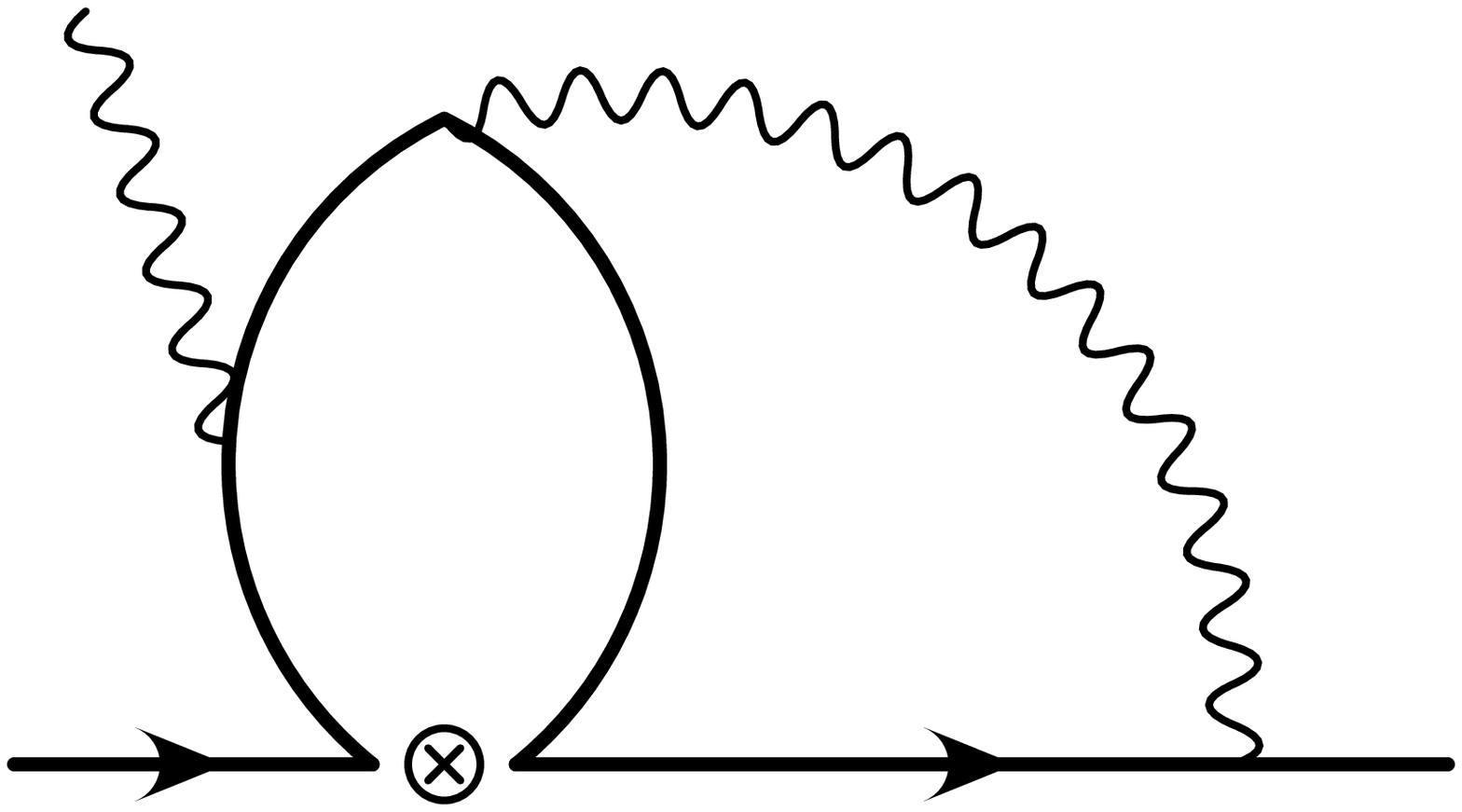,width=30mm}
&
\psfig{figure=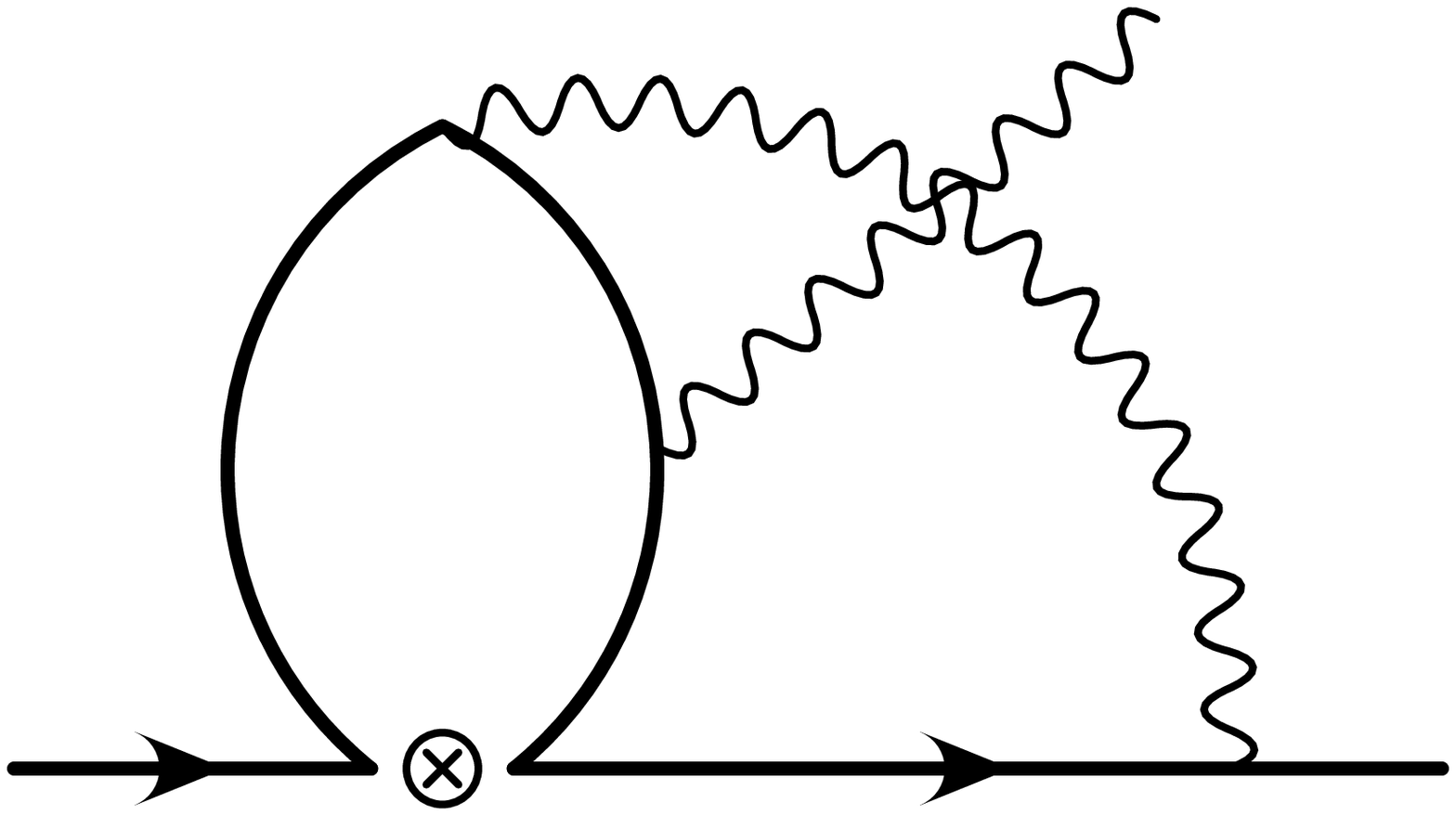,width=30mm}
&
\psfig{figure=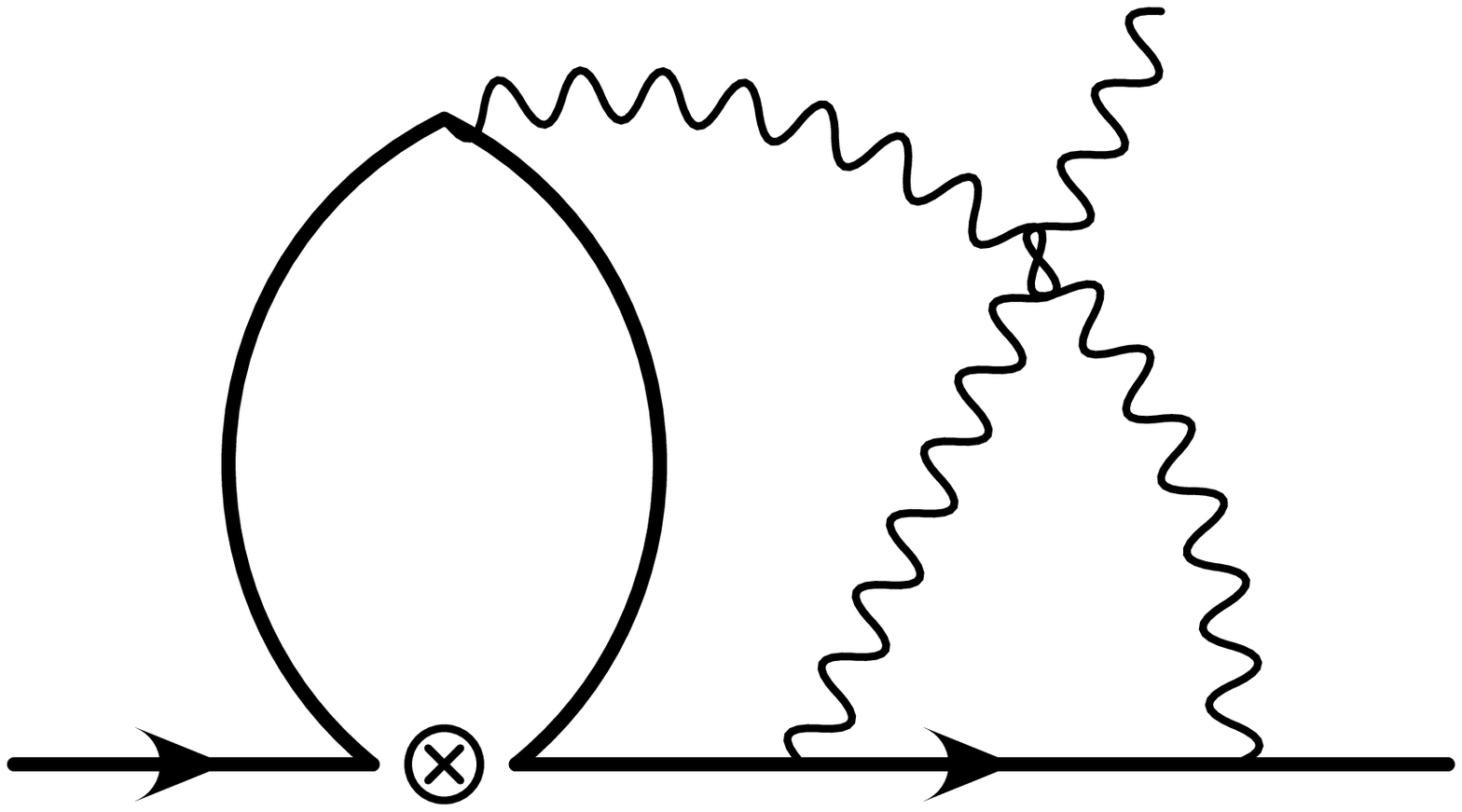,width=30mm}
\\[1mm]
 ($P_2$) & ($P_3$) & ($P_7$)
\\[3mm]
\psfig{figure=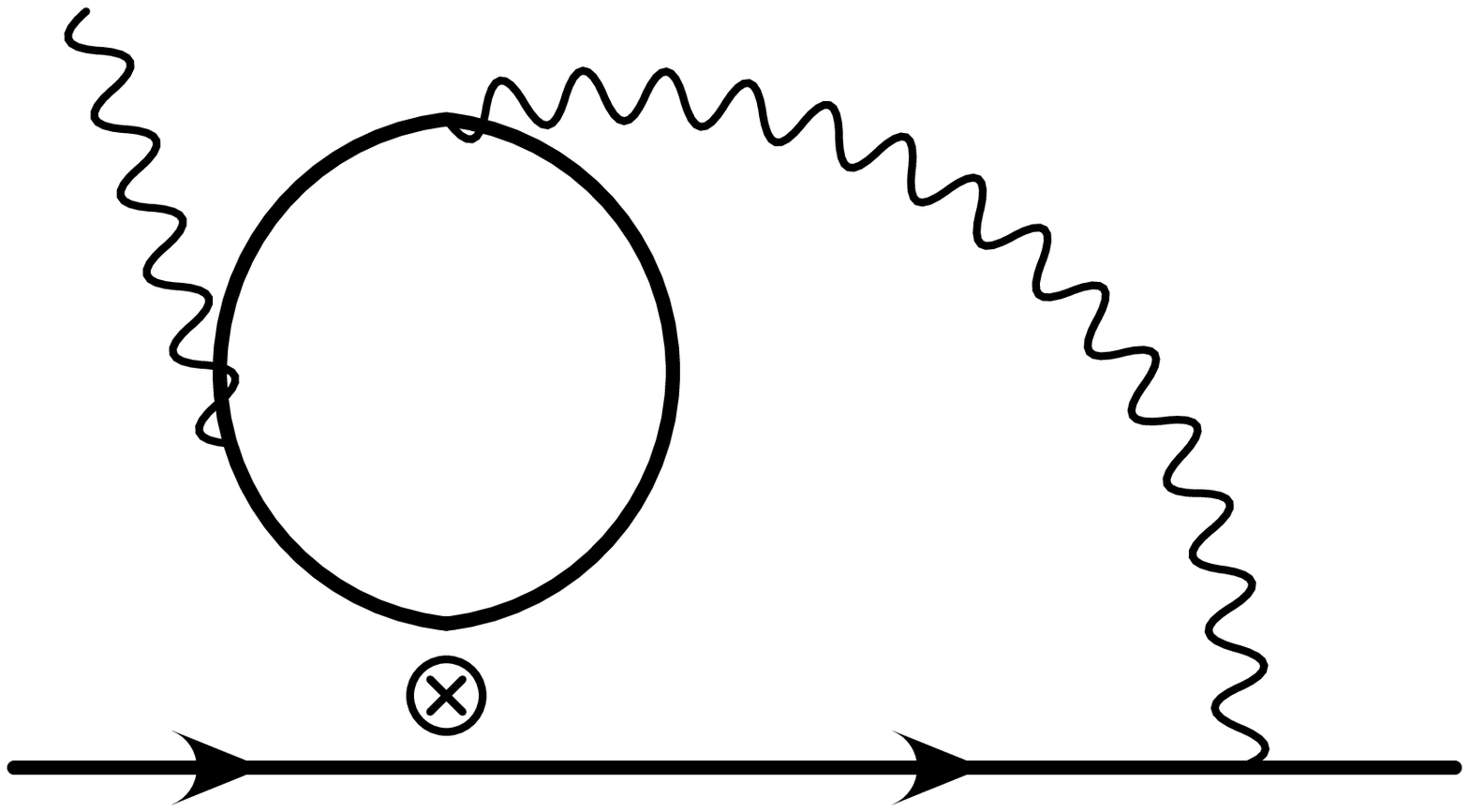,width=30mm}
&
\psfig{figure=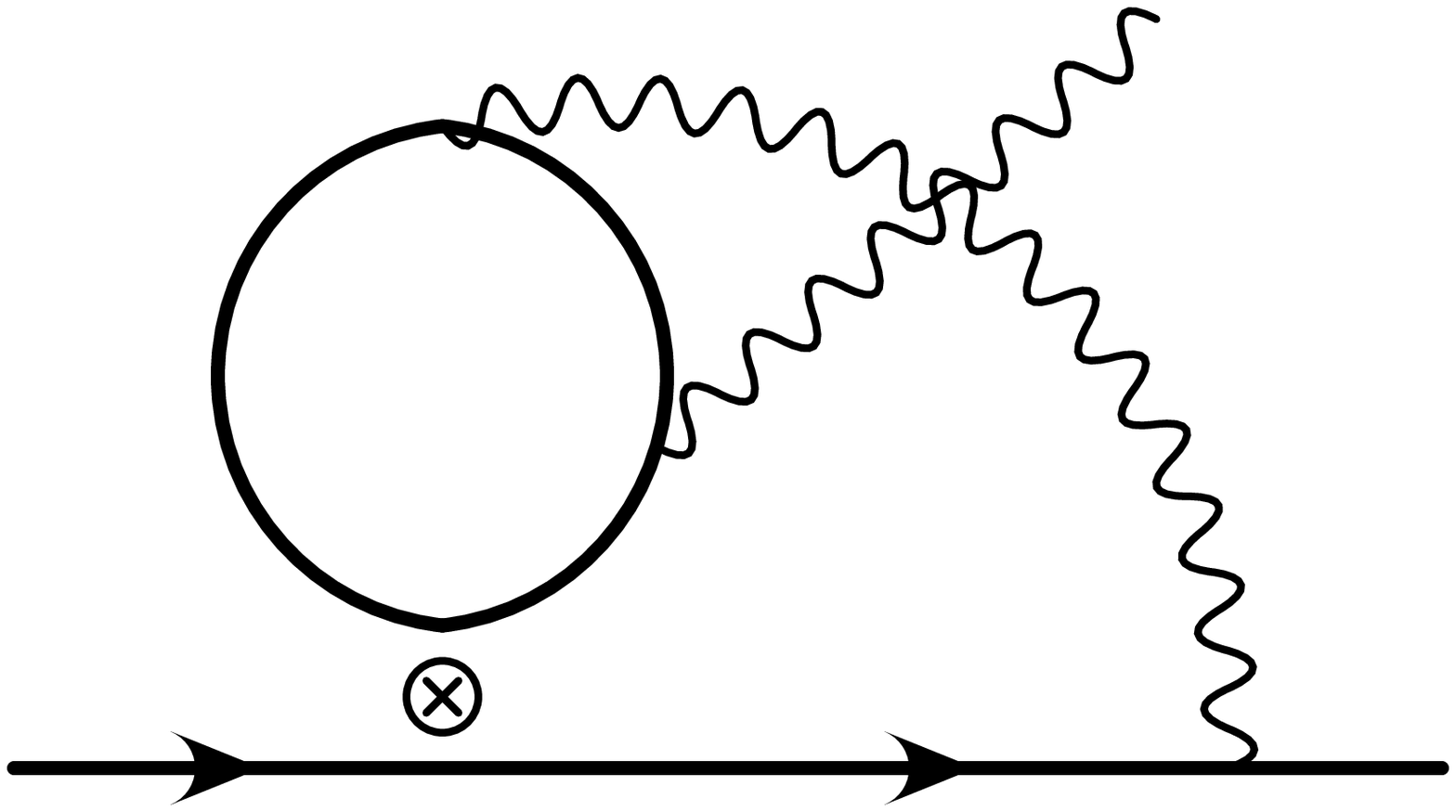,width=30mm}
&
\psfig{figure=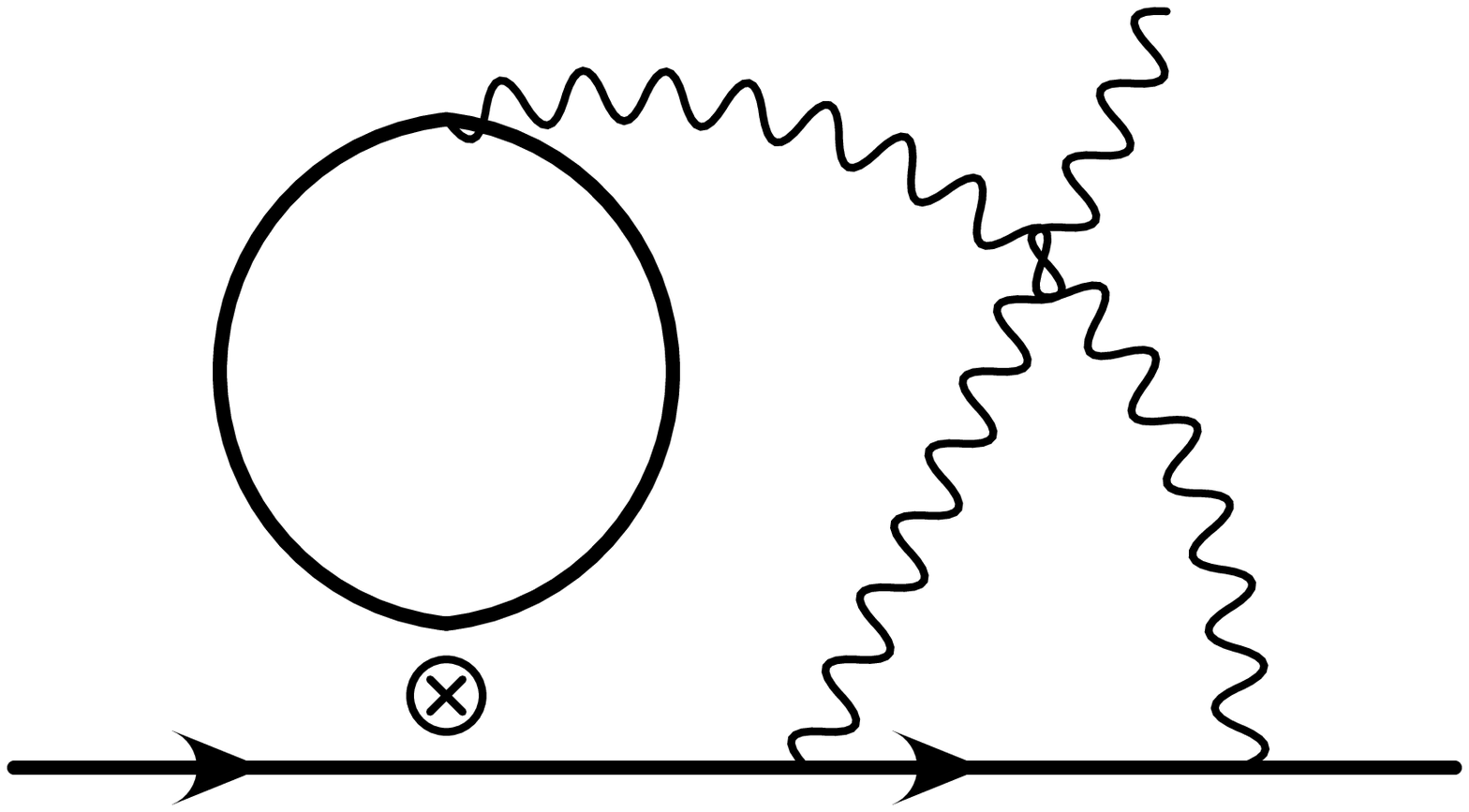,width=30mm}
\\[1mm]
 ($F_2$) & ($F_3$) & ($F_7$)
\\[1mm]
\end{tabular}
\end{minipage}
\caption{\sf \small Mixing of four-fermion operators  $V_{\mu f}$,
  $A_{\mu f}$ with $H$. The labeling of diagrams follows
Ref.~\protect\cite{Ciuchini:1994fk}.} 
\label{fig:fourmix}
\end{figure}

\begin{figure}[ht]
\hspace*{10mm}
\begin{minipage}{16.cm}
\begin{tabular}{c@{\hspace{25mm}}c}
\psfig{figure=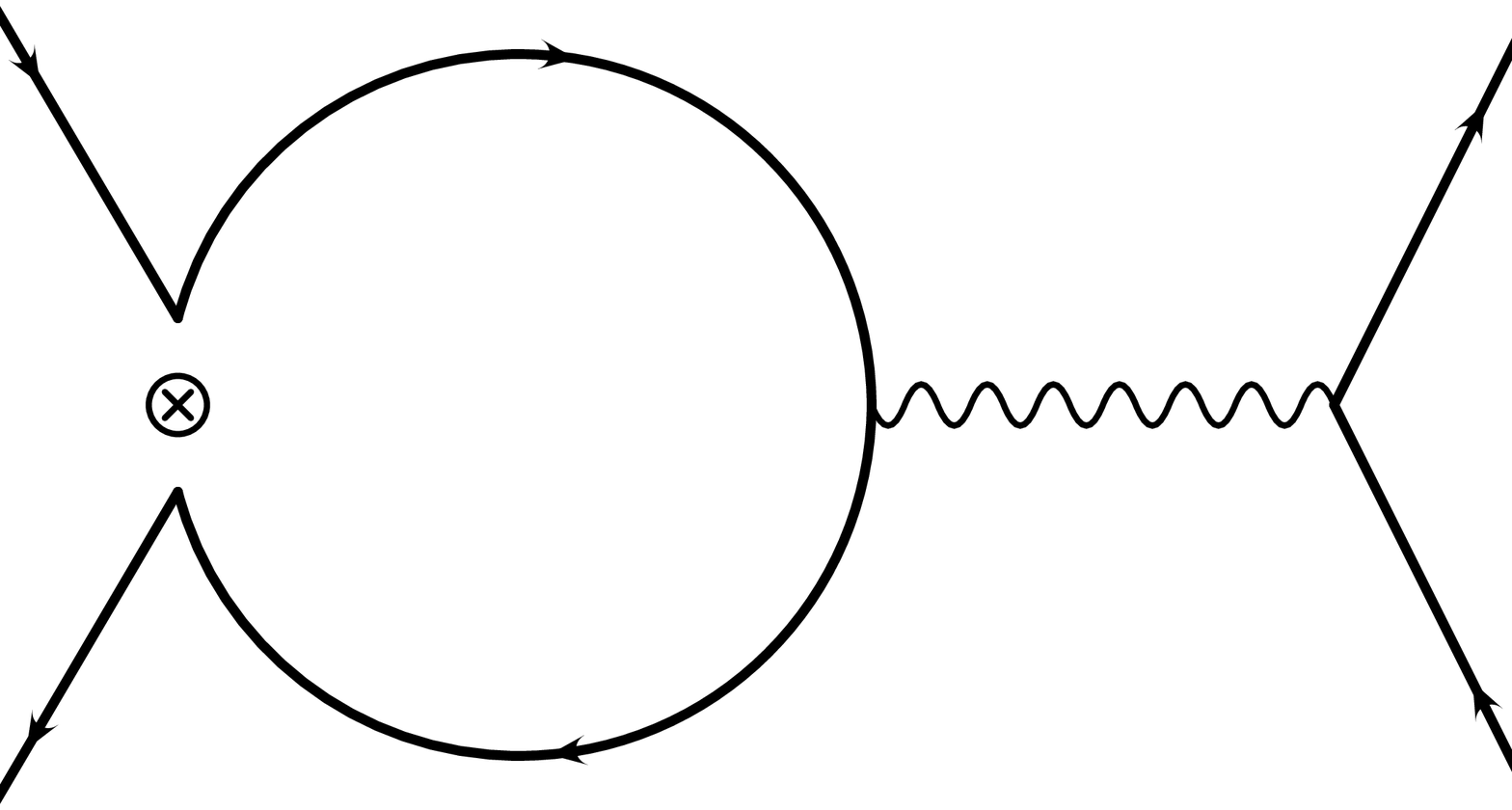,width=40mm}
&
\psfig{figure=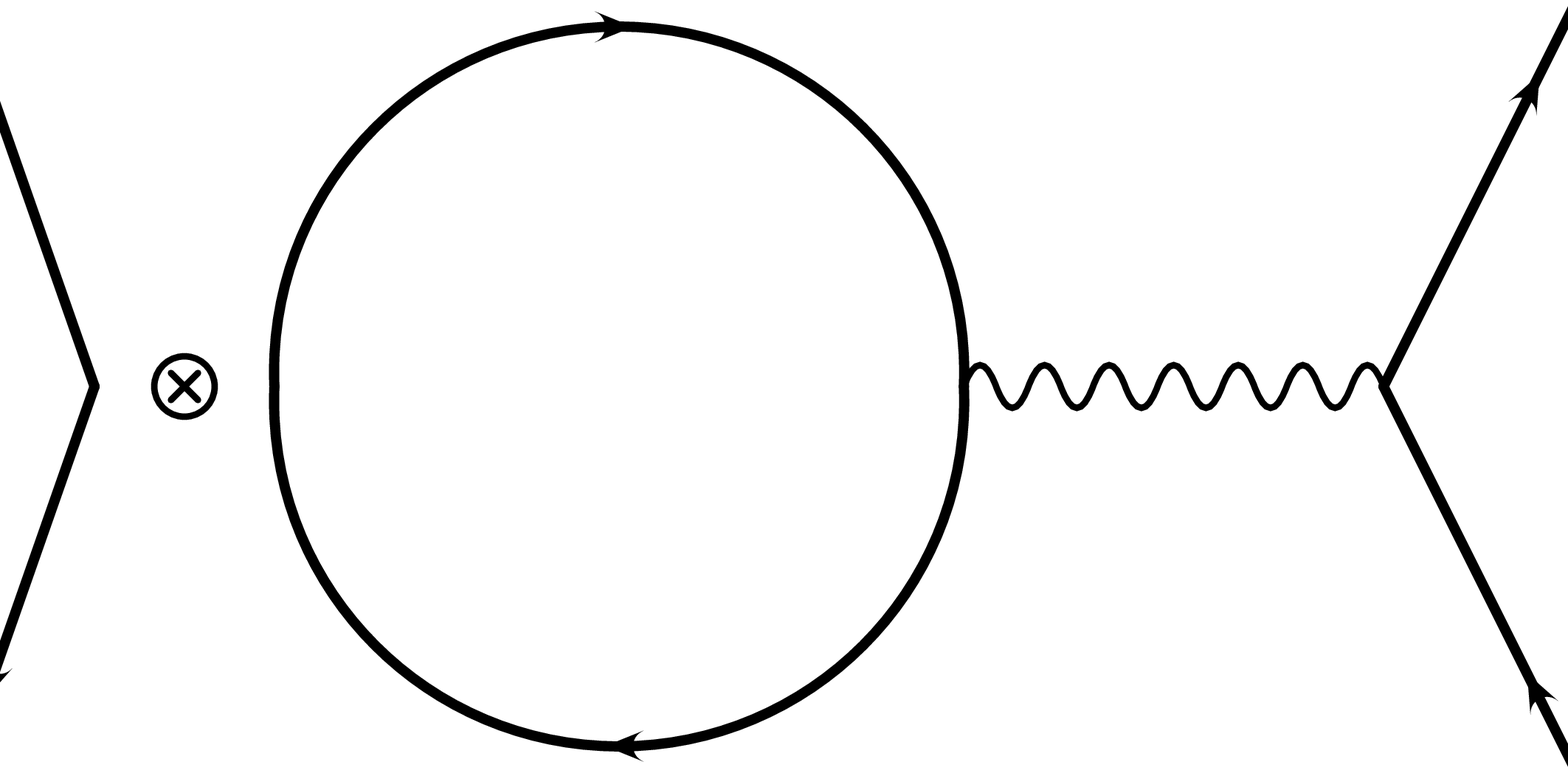,width=40mm}
\\[1mm]
 (a) & (b) 
\\[1mm]
\end{tabular}
\end{minipage}
\caption{\sf\small Renormalization of four-fermion operators: annihilation 
diagrams.}
\label{fig:4ferm34}
\end{figure}

To discuss the issues that confronted radiative quark decays, we
consider the two-loop Feynman diagrams in Fig. \ref{fig:fourmix}.
These graphs describe mixing of the four-fermion operators, designated
by $\otimes$, with the dipole operator.  Originally
\cite{Shifman:1978de} only $P_{2,3}$ and $F_{2,3}$ type diagrams were
accounted for. $P_7$ and $F_7$ were later calculated in
Refs.~\cite{Grinstein:1988vj,Grinstein:1990tj} (those authors also
accounted for one-loop mixing between gluomagnetic and electromagnetic
dipole operators).  Numerically, they did not cause much of an effect:
about $7\%$ in the mixing coefficients.  The smallness of $P_7$, $F_7$
is related to the smallness of the internal fermion loop entering as a
subgraph, which is basically a part of the photon vacuum polarization
operator (it is also non-leading in a $1/N_c$ expansion).  The
smallness of this fermion loop was used both in \cite{Shifman:1978de}
and \cite{Grinstein:1988vj,Grinstein:1990tj} to limit the number of
four-fermion operators considered to a reduced set.  The full set
(originally introduced in \cite{Shifman:1978de}) contains penguin
operators with right-handed fermions that arise from left-handed ones
due to the same fermion loop, see Fig.~\ref{fig:4ferm34}.  So, their
coefficients are also correspondingly small and could be neglected in
early studies.

The full operator basis was considered in later publications and we
refer to \cite{Ciuchini:1994fk} for a discussion of results and
references to the literature.  There, the renormalization scheme
dependence and definition of $\gamma_5$ were shown to lead to
different four-fermion operators.  As explained in
\cite{Ciuchini:1994fk}, to make the definition unambiguous, one has to
redefine the four-fermion operators by adding to them dipole operators
with appropriate coefficients.  Those coefficients are fixed by the
requirement that matrix elements of the redefined operators between
fermion and fermion plus $\gamma$ states must vanish.  In that way,
consistency among different calculational approaches was 
restored.\footnote{We disagree, however, with 
the statement in \cite{Ciuchini:1994fk} about scheme dependence in
case  of the reduced set of four-fermion operators. In our view it
is again related to the definition of four-fermion operators in 
the full set.}

Here, we note that the scheme independence can be understood in a
simpler way.  The basic point is that one-loop subgraphs for the
two-loop diagrams in Fig.~\ref{fig:fourmix} are finite and
unambiguously fixed by the use of gauge Ward identities.  A good
example is the anomalous fermion triangle involving one axial-vector
and two vector vertices where it is well known that the anomaly does
not depend on the definition of $\gamma_5$.  The logarithmic
dependence on the normalization point (scale) which comes about due to
the second loop integration is then clearly scheme independent.
Below, we use this approach, originating from
Ref.~\cite{Shifman:1978de}, to calculate all two-loop mixing.  The
results are consistent with those in Ref.~\cite{Ciuchini:1994fk} and
with the explicit two-loop leading log calculations for  $a_\mu^{\rm EW}$
given in section \ref{sec:EW}.
 These consistencies provide a useful check on the analysis.

Our calculation of the two and three loop leading logs differ,
however, from the results in \cite{Degrassi:1998es}.  The disagreement
can be traced to differences in the one and two loop anomalous
dimension matrix elements.  Details are given below.  
 
\subsection{One- and two-loop results}

As we discussed in Sec.~\ref{sec:EW} 
the electroweak contribution to the muon magnetic anomaly,
$a_\mu=(g_\mu-2)/2$, can be represented as a sum over $W$, $Z$ and
Higgs bosons.  
In one-loop the Higgs contribution is negligible and $a^{\rm EW}_\mu
\mbox{(1-loop)}$
given in Eq.\,(\ref{eq10}) is a sum of  $a_\mu^{(W)}\mbox{(1-loop)}$ and
 $a_\mu^{(Z)}\mbox{(1-loop)}$,
\begin{eqnarray}
&&a_\mu^{(W)}\mbox{(1-loop)}=\frac{G_\mu m_\mu^2}{8\pi^2\sqrt{2}}\cdot
\frac{10}{3}\;,
\nonumber\\[1mm]
&&a_\mu^{(Z)}\mbox{(1-loop)}=\frac{G_\mu m_\mu^2}{8\pi^2\sqrt{2}}\cdot 
\left[-\frac{5}{3}\,(g^\mu_A)^2+
\frac{1}{3}\,(g^\mu_V)^2\right]\,,
\label{azone}
\end{eqnarray}
where we denote the axial-vector and vector couplings of $Z$ to
the muon by $g^\mu_A$ and $g^\mu_V$. In the 
Standard Model,  
\begin{eqnarray}
g^\mu_A=2I^3_\mu=-1\,, \qquad g^\mu_V=2I^3_\mu-4Q_\mu s_W^2=4s_W^2-1\,,
\end{eqnarray}
and $g^\mu_V$ is numerically very small.

At the two-loop level  electromagnetic corrections are enhanced by
$\log m_Z/m_\mu$.  For $a_\mu^{(W)}$, the logarithmic part of the full
two-loop result \cite{CKM96} is particularly simple,
\begin{eqnarray}
a_\mu^{(W)}\mbox{(2-loop)}_{LL}=- 4\, \frac{\alpha}{\pi}
\ln\frac{m_Z}{m_\mu}\,a_\mu^{(W)}\mbox{(1-loop)}.
\label{Wcalc}
\end{eqnarray}
As we discuss below, it just reflects the anomalous dimension
of the corresponding dipole operator.
In the case of $a_\mu^{(W)}$ it is the only source of the logarithm at
two loops. 

For $a_\mu^{(Z)}$ the situation is more
complicated.
In particular, Feynman diagrams without closed fermion
loops  give \cite{KKSS,CKM96}
\begin{eqnarray}
a_\mu^{(Z)}\mbox{(2-loop; no ferm.~loops)}_{LL}
&= & 
\frac{G_\mu m_\mu^2}{8\pi^2\sqrt{2}}\cdot
\frac{\alpha}{\pi}\ln\frac{m_Z}{m_\mu}\left[\frac{13}{9}\,(g^\mu_A)^2
-\frac{23}{9}\,(g^\mu_V)^2\right]
 \nonumber\\[1mm]
=- 4\,\frac{\alpha}{\pi}
\ln\frac{m_Z}{m_\mu}\,a_\mu^{(Z)}\mbox{(1-loop)}
& + &
\frac{G_\mu m_\mu^2}{8\pi^2\sqrt{2}}\cdot
\frac{\alpha}{\pi}\ln\frac{m_Z}{m_\mu}\left[-\frac{47}{9}\,(g^\mu_A)^2
\!\!-\frac{11}{9}\,(g^\mu_V)^2\right].
\label{open}
\end{eqnarray}
In the second line we separated out the piece  due to the
anomalous dimension.

We  also have to add diagrams with closed fermion loops.
The diagrams with the muon loops give
\begin{eqnarray}
a_\mu^{(Z)}\mbox{(2-loop; muon~loops)}_{LL} = \frac{G_\mu
m_\mu^2}{8\pi^2\sqrt{2}}\cdot
\frac{\alpha}{\pi}
\ln\frac{m_Z}{m_\mu}\cdot\left[ -6N_\mu \,(g^\mu_A)^2
-\frac{4}{9}\,N_\mu\,(g^\mu_V)^2\right],
\label{closed}
\end{eqnarray}
where we introduced the factor $N_\mu $ equal to 1 for the muon loop
just to distinguish between contributions with and without closed
fermion loops. This generalizes calculations in
\cite{KKSS,Peris:1995bb,CKM95} by including the second term
proportional to  $(g^\mu_V)^2$ in
Eq.\,(\ref{closed}). This term vanishes at $s_W^2=1/4$.  In
Eq.\,(\ref{closed}) the 
first term proportional to $(g^\mu_A)^2$ arises from the induced
coupling of a $Z$ with two photons via triangle diagrams, see the
diagrams $F_2$, $F_3$ in Fig.\,\ref{fig:fourmix}. The second term,
from the vector coupling, corresponds to the $\gamma$-$Z$ mixing via a
muon loop, see the diagram $F_7$ in Fig.\,\ref{fig:fourmix}.

Fermions other than muon  contribute only via closed loops in
two-loop order.  Including their effect leads to a generalization of
Eq.\,(\ref{closed}) to
\begin{eqnarray}
\!\!\!\!\!\!&&a_\mu^{(Z)}\mbox{(2-loop; ferm.~loops)}_{LL}
\nonumber\\[1mm]
\!\!\!\!\!\!&&
=\frac{G_\mu m_\mu^2}{8\pi^2\sqrt{2}}\cdot
\frac{\alpha}{\pi}
\,\sum_f
\ln\frac{m_Z}{
\{m_f,m_\mu\}}
\left[ -6 \,g^\mu_A g_A^f N_f Q_f^2
+\frac{4}{9}\,g^\mu_V g_V^f N_f Q_f\right],
\label{allclosed}
\end{eqnarray}
where we introduced the notation
\begin{eqnarray}
\{m_f, m_\mu\} \equiv
{\rm max}\{m_f,m_\mu\}\,.
\label{mmass}
\end{eqnarray}
Moreover, 
$Q_f$ is the electric charge of the fermion, $N_f=1$  for
leptons and $N_f=N_c=3$ for quarks, and
\begin{eqnarray}
g_A^f=2I^3_f\,,\qquad g_V^f=2I^3_f-4s_W^2 Q_f\,.
\label{gAV}
\end{eqnarray} 
It is implied that $m_f\ll m_Z$ in Eq.\,(\ref{allclosed}); so, it does not
include the top quark contribution which is part of
the nonlogarithmic, NLL, terms.

The closed loop contribution (\ref{allclosed}) and the last term in the
second line of Eq.\,(\ref{open}) are due to the two-loop mixing with
four-fermion operators to be discussed below.  Overall, the two-loop
result for the sum of $a_\mu^{(W)}$ and $a_\mu^{(Z)}$ is (the form is
slightly different than Eq.\,(\ref{eq17}) in Sec.~\ref{sec:EW}, but
equivalent) 
\begin{eqnarray}
a_\mu^{(W,Z)}\mbox{(2-loop)}_{LL}&=&\frac{G_\mu
m_\mu^2}{8\pi^2\sqrt{2}}\cdot
\frac{\alpha}{\pi}
\left\{ -\left[\frac{215}{9}+\frac{31}{9}\,(g^\mu_V)^2\right]
\ln\frac{m_Z}{m_\mu}\right.
\nonumber\\[1mm]
&&+ \left. \sum_{f=u,d,s,c,\tau ,b}
\left[ 6\,g_A^fN_f Q_f^2
+\frac{4}{9}\,g^\mu_V g_V^fN_f Q_f\right]\ln\frac{m_Z}{m_f}
\right\},
\label{total2}
\end{eqnarray}
where we neglected the mass difference between $W$ and $Z$
(the $\ln(m_Z/m_W)$ terms  are put into NLL 
contributions). The first
term in  Eq.\,(\ref{total2}) accounts for diagrams with muons and electrons
and in the second
term the sum is over all other fermions except
top. The dependence on $s_W^2$ enters
via $g_V^\mu$, $g_V^f\,$.  Our Eq.\,(\ref{total2}) differs somewhat from
Eq.\,(25) in \cite{Degrassi:1998es}, as explained at the end of section
\ref{sub:anom}.

\subsection{Effective Lagrangian}

In the effective Lagrangian, represented as a sum over local operators
normalized at a point $\mu$, the anomalous magnetic moment of the
fermion $f$ is associated with the operator $F_{\alpha\beta} \, \bar f
\,\sigma^{\alpha\beta} f$ of dimension 5.  Because of a chirality flip
it enters with a coefficient proportional to the fermion mass $m_f$,
and it is convenient to include $m_f$ in the definition of the
operator,
\begin{eqnarray}
H(\mu)= -{m_\mu(\mu)\over 16\pi^2} \left[eF_{\alpha\beta}\,\bar \mu\,
\sigma^{\alpha\beta}\mu\right]_\mu
\,, 
\label{dipop}
\end{eqnarray}
where the electric charge $e= \sqrt{4\pi\alpha}$ is another factor
included in the operator definition. Both the mass $m_\mu$ and the
electric charge $e$ are $\mu$ dependent quantities in
Eq.\,(\ref{dipop}) but the running of the electric charge $e$ is
canceled by the wave function renormalization of the electromagnetic
strength tensor $F_{\alpha\beta}$.  The product $eF_{\alpha\beta}$ is
RG invariant.

The effective Lagrangian for flavor and parity preserving transitions
can be written as 
\begin{eqnarray}
{\cal L}_{\rm eff}(\mu)=-\frac{G_\mu}{2\sqrt{2}} \left\{ h(\mu)
H(\mu) + \sum_i 
c^i(\mu){\cal O}_i(\mu)\right\}\,,
\end{eqnarray}
where the second sum extends over $d=6$ four-fermion 
operators.\footnote{We omit here the $d=5$ dipole operators for other
fermions as well as chromomagnetic dipole operators for quarks, since they
do not mix with $H\,$.}  The observable value of $a^{\rm EW}_\mu$ is
related to the coefficient $h$ at the low normalization point
$\mu=m_\mu\,$,
\begin{eqnarray}
a^{\rm EW}_\mu= \frac{G_\mu\,m_\mu^2}{8\pi^2\sqrt{2}}\, h(m_\mu)
\label{aZ}
\end{eqnarray}
The relation (\ref{aZ}) implies that only the operator $H$ contributes
to the matrix elements $\langle \mu |{\cal L}_{\rm
eff}(m_\mu)|\mu \gamma\rangle$ --- a condition which we discussed above.
 
The one-loop results (\ref{azone}) refer to the range of virtual momenta
of order $m_W$ and $m_Z$, thus fixing the value of $h$ at the
high normalization point,
\begin{eqnarray}
 h^{(W)}(m_W)=\frac{10}{3}\,, \qquad   
h^{(Z)}(m_Z)=-\frac{5}{3}\,(g^\mu_A)^2+
\frac{1}{3}\,(g^\mu_V)^2\,.
\label{onelc}
\end{eqnarray} 

We choose the basis for the four-fermion operators defining them as
\begin{eqnarray}
{\cal O}_{V;f g}=\frac{1}{2}\,\bar f\gamma^\nu f\, \bar g\gamma_\nu
g\,,\qquad 
{\cal O}_{A;f g}=\frac{1}{2}\,\bar f\gamma^\nu\gamma_5 f\,
\bar g\gamma_\nu\gamma_5 g 
\label{VAop}
\end{eqnarray}
The $d=6$ part of the effective Lagrangian can be written as
\begin{eqnarray}
{\cal L}^{d=6}(\mu) =-\frac{G_\mu}{2\sqrt{2}}
\sum_{f,g,\Gamma=V,A} c^{\Gamma;fg}(\mu) \,{\cal O}_{\Gamma;fg}\,,
\qquad c^{\Gamma;fg}(\mu) \equiv c_{_{(Z)}}^{\Gamma;fg}(\mu) 
 + c_{_{(W)}}^{\Gamma;fg}(\mu),
\label{d6}
\end{eqnarray}
where ${\cal O}_{\Gamma;fg}={\cal O}_{\Gamma;gf}$; so, the OPE
coefficients $c^{\Gamma;fg}$ are symmetric under permutation of $f$
and $g\,$.  The tree-level $Z$ exchange gives the initial data for the
OPE coefficients,
\begin{eqnarray}
c_{_{(Z)}}^{\Gamma;fg}(m_Z)=g^f_{_\Gamma}\,  g^g_{_\Gamma} \,,\qquad
(\Gamma =V,A)\,,
\label{initVA}
\end{eqnarray}
where the vector and axial-vector couplings $g^f_{V,A}$ are given in
Eq.\,(\ref{gAV}).

\begin{figure}[ht]
\hspace*{0mm}
\begin{minipage}{16.cm}
\psfig{figure=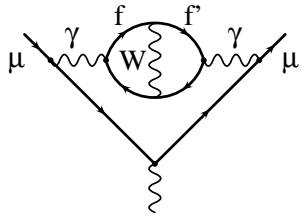,width=40mm}
\end{minipage}
\caption{\sf \small Contribution to $a^{\rm EW}_\mu$
from operators in ${\cal L}_W^{d=6}(m_W)$. Both fermions $f$ and $f'$
must be charged so leptons do not contribute.}
\label{fig:gW}
\end{figure}

In the case of $W$ exchange  the tree-level effective Lagrangian is
\begin{eqnarray}
{\cal L}_W^{d=6}(m_W) = -\frac{G_\mu}{\sqrt{2}}\,
\left[ \bar u\gamma^\nu(1-\gamma _5) d \,\bar d\gamma_\nu(1-\gamma _5)u
+(u\to c,~d\to s)\right],
\label{LW} 
\end{eqnarray}
where we neglect CKM mixing.  Such operators contribute to $h$ only if
all fermions are charged, so we can omit the leptonic part (see
Fig.~\ref{fig:gW}).  One can use the Fierz transformation to put the
operator in the following form:
\begin{eqnarray}
\bar u\gamma^\nu(1-\gamma _5) d \,\bar d\gamma_\nu(1-\gamma _5)u&=&
\frac{1}{N_c}\,\bar u\gamma^\nu(1-\gamma _5) u \,\bar d\gamma_\nu(1-\gamma
_5)d \nonumber\\[1mm]
&&+ 2\,\bar u\, t^a\gamma^\nu(1-\gamma _5) u \,\bar d\,
t^a\gamma_\nu(1-\gamma _5)d\,,
\end{eqnarray}
where $t^a$ are matrices of the color generators. The part with $t^a$ does
not contribute to the magnetic moment (up to gluon corrections which
we do not consider here) and neither does the parity breaking part, so
${\cal L}_W^{d=6}$ reduces to the form (\ref{d6}) with the initial data
\begin{eqnarray}
c_{_{(W)}}^{\Gamma;ud}(m_W)=c_{_{(W)}}^{\Gamma;du}(m_W)=\frac{1}{N_c}\,,
\qquad
(\Gamma =V,A)\,.
\label{initW}
\end{eqnarray}
There are similar operators and coefficients for $u,d$ substituted by
$c,s$.  These are the only operators that contribute to three loop
leading log mixing in the $W$ sector.

The RG equations which allow us to calculate the running of $h(\mu)$ are
\begin{eqnarray}
&&\mu\,\frac{{\rm d} h(\mu)}{{\rm d} \mu}=-\frac{\alpha(\mu)}{2\pi}
\left[\gamma_{_H} h(\mu) +\sum_{\Gamma,f,g}\beta_{\,\Gamma;fg}
\,c^{\Gamma;fg}(\mu)\theta(\mu-m_{fg})
\right],
\label{RGeq1}
\\[1mm]
&&\mu\,\frac{{\rm d} c^{\Gamma;fg}(\mu)}{{\rm d}
\mu}=-\frac{\alpha(\mu)}{2\pi}\sum_{\Gamma',f',g'}
\gamma^{^{\Gamma; fg }}_{_{\Gamma'
;f'g'}}\,c^{\Gamma';f'g'}(\mu)\theta(\mu-m_{f'g'})\,,
\label{RGeq2}
\\[1mm]
&&\mu\,\frac{{\rm d} \alpha(\mu)}{{\rm d} \mu}=-\frac{\alpha^2(\mu)}{2\pi}\,
\sum_f b_f \theta(\mu-m_f)\,,
\quad
b_f=-\frac{4}{3}\, N_f Q_f^2\,.
\label{RGeq3}
\end{eqnarray}
Here $\gamma_{H}$, $\beta_{\Gamma;fg}$ and $\gamma^{\Gamma;
fg}_{\Gamma';f'g'}$ form the matrix of anomalous dimensions for
operators $H$ and ${\cal O}_{\Gamma;fg}\,$.  That matrix is
``block--triangular'': $H$ does not mix with the $d=6$ operators but
the operators ${\cal O}_{fg}^\Gamma$ do mix with $H$.  The
$\beta_{\Gamma;fg}$ correspond to these mixings.\footnote{Note that
our definition of anomalous dimensions differs from that in
Refs.~\cite{Ciuchini:1994fk,Degrassi:1998es} by a factor
$(-1/2)\,$. Also the normalization of four-fermion operators with
$f\neq g$ is different.}  The $\theta$ functions in r.h.s.\ count only
active fermions at the given $\mu$, with $m_{fg}$ denoting the maximal
fermion mass in the operator ${\cal O}_{fg}$\,,
\begin{eqnarray}
m_{fg}=\{m_f,m_g\}\equiv {\rm max}\, \{m_f,m_g\}\,.
\end{eqnarray}

Perturbatively $h=h^{(1)}+h^{(2)}+h^{(3)}+\ldots$ where the raised index
denotes the number of loops. In one-loop approximation
$h^{(1)}(\mu)=h(M)$ where $h(M)$ is given in Eq.\,(\ref{onelc}). In
two-loop order one can neglect in the r.h.s.\ of Eq.\,(\ref{RGeq1}) the
running of $\alpha$, $c_H$ and $c^{\Gamma ; fg}$, and get
\begin{eqnarray}
h^{(2)}(\mu)=\frac{\alpha(M)}{2\pi}\left[\gamma_{_H} h(M)\ln\frac{M}{\mu} +
\sum_{\Gamma,f,g}\beta_{\,\Gamma;fg}
\,c^{\Gamma;fg}(M)\ln\frac{M}{\{\mu,m_{fg}\}}
\right]\,.
\label{2loopc}
\end{eqnarray}
Below we will compute $\gamma_{H} $ and $\beta_{\Gamma;fg}$ and then
verify as a check
that Eq.\,(\ref{2loopc}) matches the explicit two-loop calculations. We
will then use the RG equations (\ref{RGeq1}\,--\ref{RGeq3}) to
determine $h$ in three loops.

\subsection{Anomalous dimensions and mixing of effective operators}
\label{sub:anom}

\begin{figure}[ht]
\hspace*{0mm}
\begin{minipage}{16.cm}
\begin{tabular}{c@{\hspace{16mm}}c@{\hspace{16mm}}c}
\psfig{figure=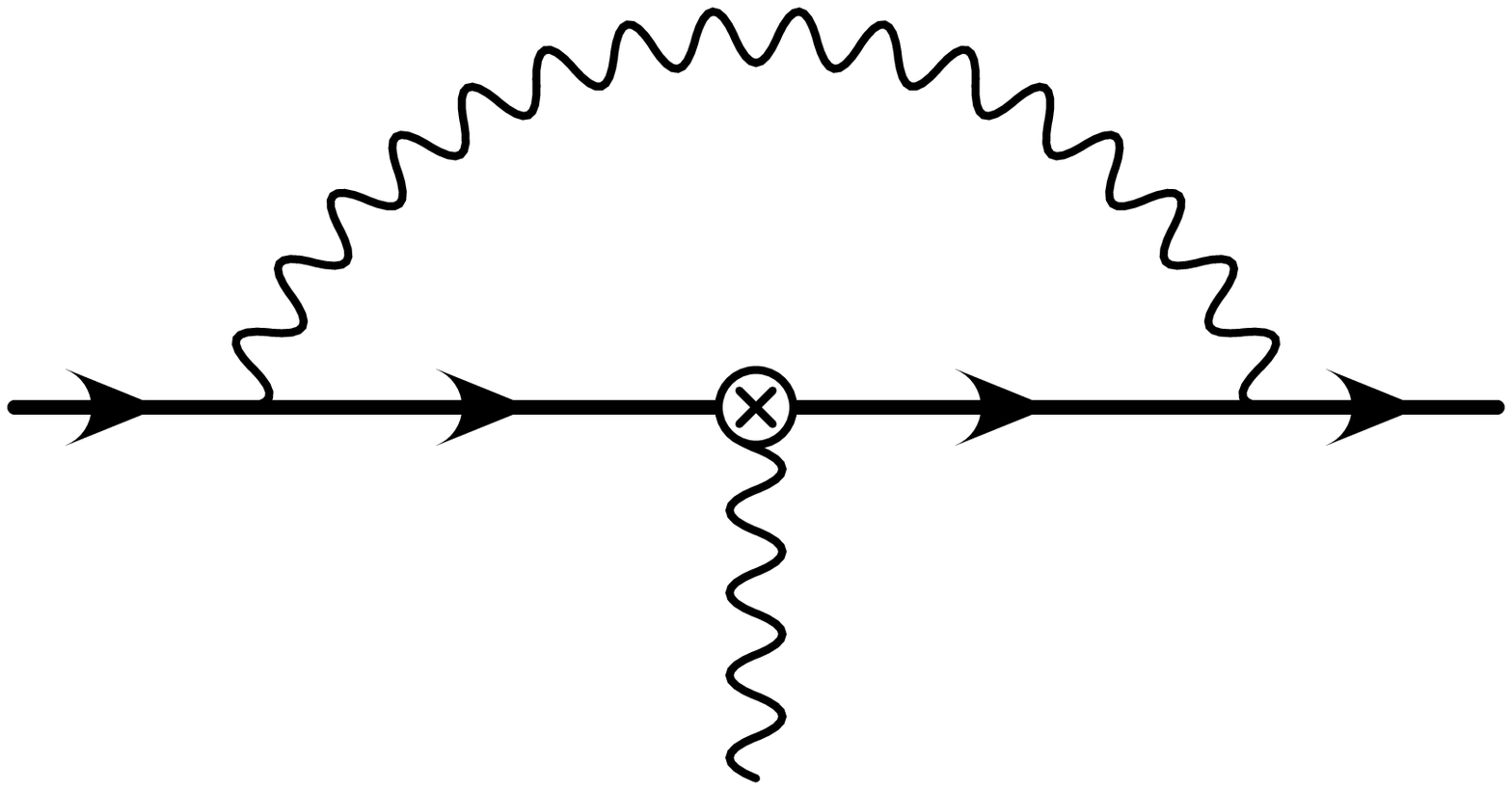,width=30mm}
&
\psfig{figure=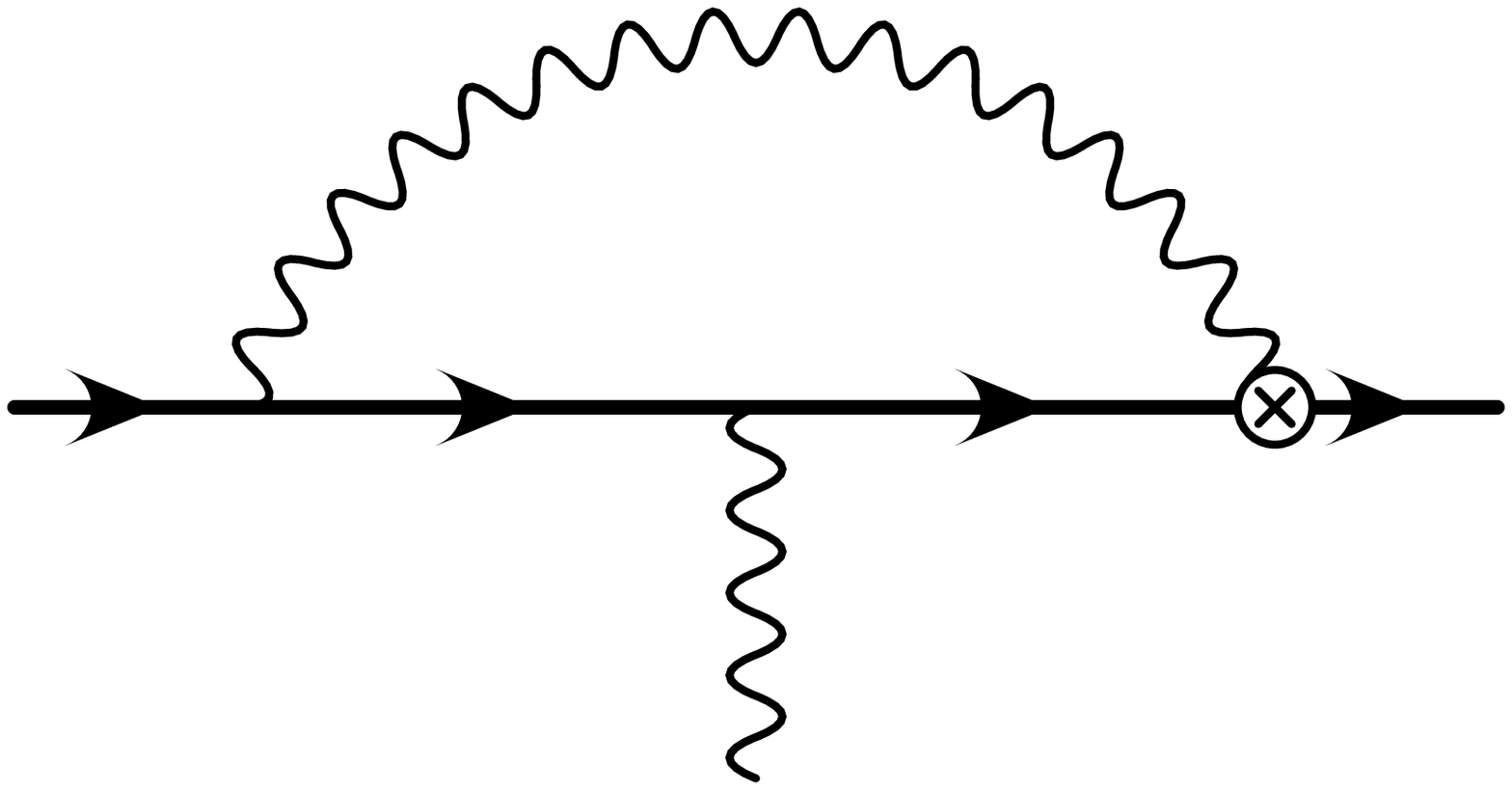,width=30mm}
&
\psfig{figure=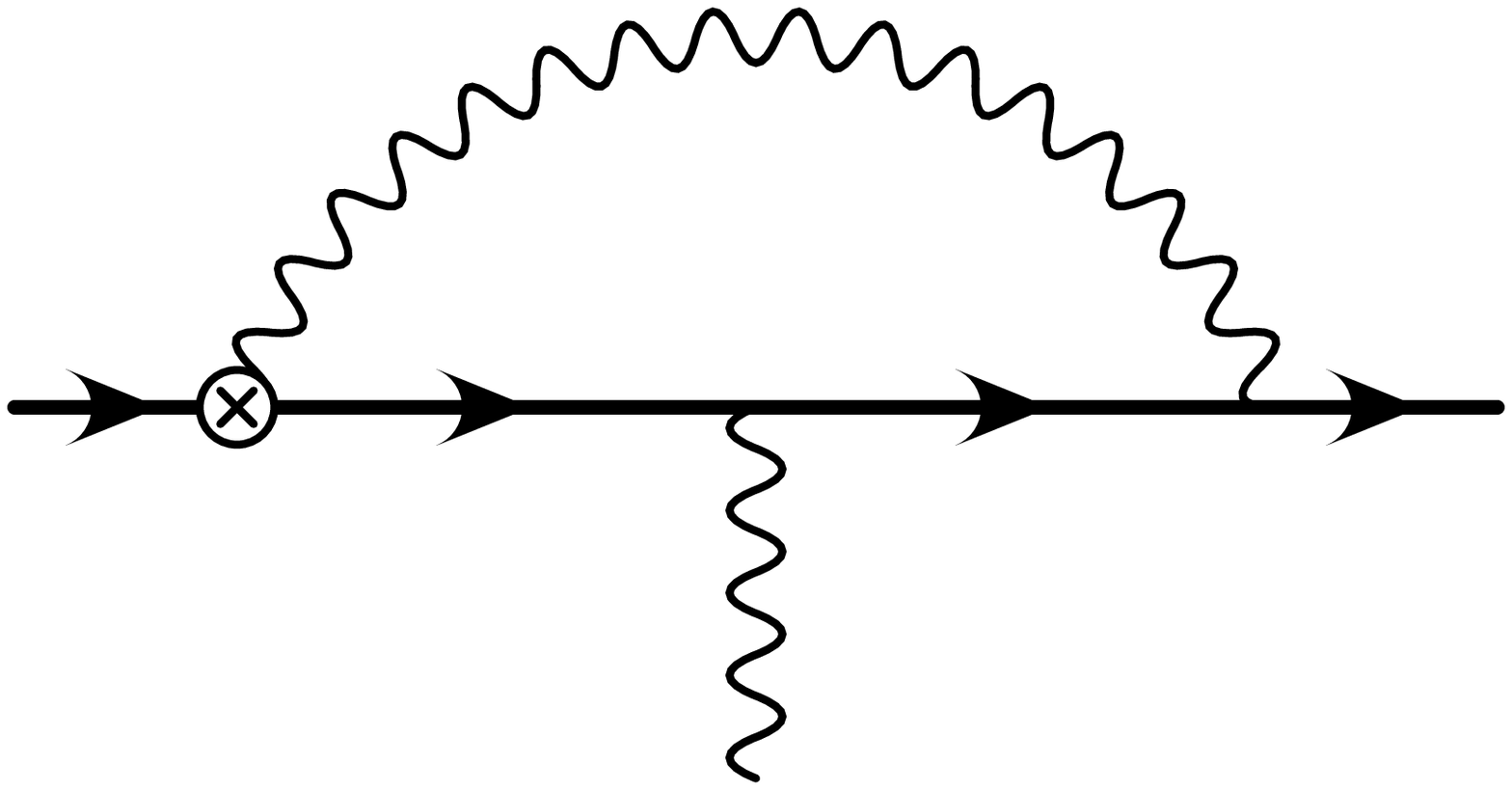,width=30mm}
\\[1mm]
 (a) & (b) & (c)
\\[1mm]
\end{tabular}
\end{minipage}
\caption{\sf \small Anomalous dimension of the dipole operator $H$.}
\label{fig:dipol}
\end{figure}

First, consider the anomalous dimension $\gamma_{H}$ of the dipole
operator $H$.  It is conveniently computed in the Landau gauge, with a
photon propagator $-i(g_{\mu\nu}-k_\mu k_\nu/k^2)/k^2$, since in this
gauge there are no logs in $Z$ factors. Thus, $\gamma_{H}$ is given by
the sum of three diagrams in Fig.\,\ref{fig:dipol} 
minus anomalous
dimension $\gamma_m=3$ of the mass $m_\mu(\mu)$ included into the
definition (\ref{dipop}) of $H$.
The result is
\begin{eqnarray}
\gamma_{_H}=-1-2-2-3=-8\,,
\label{anomdim}
\end{eqnarray}
where the numbers correspond to diagrams $a$, $b$, $c$ and
$(-\gamma_m)$.

More involved two-loop calculations are needed to determine the
mixings $\beta_{\,\Gamma;fg}$ of the four-operators (\ref{VAop}) with
$H$.  The relevant diagrams are shown in Fig.\,\ref{fig:fourmix}.  It
is clear that the operator $O_{\Gamma;fg}$ mixes with $H$ only when at
least one of its fermionic indices coincides with the muonic one.

Let us start with operator ${\cal O}_{V;\mu f}$ with $f\neq \mu
\,$. It is the diagram $F_7$ (plus, of course, a similar diagram where
the virtual photon is coupled to the incoming muon leg) which defines
$\beta_{V;\mu f}\,$. The fermion loop in this diagram (which is the
same as in the photon polarization operator) produces 
\begin{eqnarray}
eN_f Q_f\, \frac{Q^2}{12\pi^2}\,\ln\frac{M^2}{Q^2}\,,
\label{polar}
\end{eqnarray}
where $Q$ is the Euclidean momentum of the virtual photon. The $Q^2$
factor in (\ref{polar}) cancels the photon propagator and for the second
loop integration we get an expression similar to the one-loop $Z$
boson exchange with the pure vector coupling, $g_{\,V}=1$, $ g_{A}=0$,
up to the substitution
\begin{eqnarray}
\frac{m_Z^2}{m_Z^2+Q^2}\Longrightarrow \frac{e^2N_f Q_f}{24\pi^2}\,
\ln\frac{M^2}{Q^2}\,.
\end{eqnarray}
The $\ln(M^2/Q^2)$ can be represented as
\begin{eqnarray}
\ln\frac{M^2}{Q^2}=\int_{\mu^2}^{M^2}\frac{{\rm d} \widetilde M^2}
{ \widetilde M^2+Q^2}
\end{eqnarray}
with the range $\mu^2\ll Q^2\ll M^2$. That allows us to get the two-loop
result from the one-loop one. From Eq.\,(\ref{onelc}) we see that
$h^{(Z)}=1/3$ at $g_{\,V}=1$, $g_{A}=0$. Thus, the diagram $F_7$ produce
\begin{eqnarray}
\frac{e^2N_f Q_f}{24\pi^2}\cdot \frac{1}{3}\int_{\mu^2}^{M^2}
 \frac{{\rm d} \widetilde M^2}
{ \widetilde M^2}=\frac{2}{9}\cdot \frac{e^2N_f Q_f}{16\pi^2}\,
\ln\frac{M^2}{\mu^2}\,.
\end{eqnarray}
in $h^{(Z)}\,$.
 It gives for $\beta_{V;\mu f}$
\vspace{-4mm}
\begin{eqnarray}
\beta_{V;\mu f}=\frac{4}{9}\,N_f Q_f\,,\qquad f\neq\mu\,,
\label{bVf}
\end{eqnarray}
where we accounted for another diagram similar to $F_7\,$.

For the operator ${\cal O}_{A;\mu f}$ with $f\neq \mu $ its mixing
with $H$ is given by the diagrams $F_2$ and $F_3$. The fermion loop in
this case is the anomalous triangle with one axial-vector and two
vector vertices. We need its kinematics when the momentum of the
external photon tends to zero. The triangle then reduces to
(see Eqs.\,(\ref{invfun}) and (\ref{msquare})) 
\begin{eqnarray}
\frac{e^2 N_f Q_f^2}{2\pi^2}\left[\tilde F_{\mu\nu}-
\frac{q_\mu q_\sigma}{q^2}\,\tilde F_{\sigma\nu}-
\frac{q_\nu q_\sigma}{q^2}\,\tilde F_{\sigma\mu}\right],
\end{eqnarray}
where $q$ is the momentum of virtual photon and $\tilde
F_{\mu\nu}=(1/2)\epsilon_{\mu\nu\sigma\delta}F^{\sigma\delta}$ is the
dual of the external electromagnetic field $F_{\mu\nu}$. Integration
over $q$ in the second loop is logarithmic and produces 
\begin{eqnarray}
\beta_{A;\mu f}=-6\,N_f Q_f^2\,,\qquad f\neq\mu\,.
\label{bAf}
\end{eqnarray} 

For the pure muonic operators ${\cal O}_{\Gamma;\mu\mu } $ their
mixing with $H$ results from diagrams $F_{2,3,7}$ as well as
diagrams $P_{2,3,7}$ in Fig.\,\ref{fig:fourmix}. The $F$ diagrams which
are due to pairing of fermions from the same current in the
four-fermion operators coincide (up to substitutions $N_f\to N_\mu=1$,
$Q_f \to Q_\mu=-1$ and the combinatorial factor two due to two ways of
pairing) with those of the ${\cal O}_{\Gamma ;\mu f}$
operators, discussed above.

We can transform the $P$ diagrams, which are due to pairing of
fermions from different currents, into $F$-type diagrams with closed
fermionic loops using Fierz transformations,
\begin{eqnarray}
\bar \psi_1 \gamma^\nu \psi_2 \, \bar \psi_3 \gamma_\nu \psi_4 &=&
\frac{1}{2}\,\bar \psi_1 \gamma^\nu \psi_4 \, \bar\psi_3 \gamma_\nu
\psi_2+
\frac{1}{2}\,\bar \psi_1 \gamma^\nu\gamma_5 \psi_4 \, \bar\psi_3 \gamma_\nu
\gamma_5\psi_2\nonumber\\[1mm]
&&-
\bar \psi_1 \psi_4 \, \bar\psi_3 \psi_2+
\bar \psi_1 \gamma_5 \psi_4 \, \bar\psi_3
\gamma_5\psi_2\,,\nonumber\\[1mm]
\bar \psi_1 \gamma^\nu\gamma_5 \psi_2 \,
\bar\psi_3 \gamma_\nu\gamma_5\psi_4 &=&
\frac{1}{2}\,\bar \psi_1 \gamma^\nu \psi_4 \, \bar\psi_3 \gamma_\nu
\psi_2+
\frac{1}{2}\,\bar \psi_1 \gamma^\nu\gamma_5 \psi_4 \, \bar\psi_3 \gamma_\nu
\gamma_5\psi_2\nonumber\\[1mm]
&&+
\bar \psi_1 \psi_4 \, \bar\psi_3 \psi_2-
\bar \psi_1 \gamma_5 \psi_4 \, \bar\psi_3 \gamma_5\psi_2\,.
\label{fierz}
\end{eqnarray}
We see that the $P$ diagrams can be reduced to already calculated $F$
ones (first two terms in the r.h.s.\ of Eqs.\,(\ref{fierz})) and to
diagrams of the $F_{2,3}$ type where instead of products of
axial-vector currents in the four-fermion operators we have scalar or
pseudoscalar ones.

Taken separately, the fermion triangles with the scalar and
pseudoscalar vertices contain logs and produce double logs in the
anomalous magnetic moment. But for the difference of scalar and
pseudoscalar operators entering Eqs.\,(\ref{fierz}) these double log terms
cancel. What remains in this combination can be presented as a piece
in the pseudoscalar triangle of the form
\begin{eqnarray}
-\frac{m_\mu}{2\pi^2 q^2}\, q_\sigma \tilde F_{\sigma\mu}\,.
\end{eqnarray}
The second loop integration is then simple. 

Altogether, it results in the following mixings of ${\cal
O}_{\Gamma;\mu \mu} $ with $H$
\begin{eqnarray}
&&\beta_{V;\mu \mu}=-\frac{4}{9}-6+4 
+2N_\mu\left(-\frac{4}{9}\right)=-\frac{22}{9}-\frac{8}{9}\,N_\mu\,,
\nonumber\\[1mm]
&&\beta_{A;\mu \mu}=-\frac{4}{9}-6-4 
+2N_\mu\,\left(-6\right)~=-\frac{94}{9}-12 N_\mu\,.
\label{bVA}
\end{eqnarray}   
The numbers written after the first equality signs display a
decomposition in terms of closed loops: vector, axial-vector and
scalar plus pseudoscalar loops. The latter piece has different signs
for $\beta_{V;\mu \mu}$ and $\beta_{A;\mu \mu}\,$.  We can unify the
expressions (\ref{bVf}), (\ref{bAf}) and (\ref{bVA}) as
\begin{eqnarray}
&&\beta_{V;fg}=\delta^\mu _f\cdot\frac{4}{9}\,N_g Q_g+\delta^\mu
_g\cdot\frac{4}{9}\,N_f Q_f+\delta^\mu _f\,\delta^\mu
_g\cdot\left(-\frac{22}{9}\right),
\nonumber\\[1mm]
&&\beta_{A;fg}=\delta^\mu _f\cdot(-6N_g Q^2_g)+\delta^\mu
_g\cdot(-6N_f Q^2_f)+\delta^\mu _f\,\delta^\mu _g
\cdot\left(-\frac{94}{9}\right).
\label{genbVA}
\end{eqnarray}

Now we are well prepared to compare the RG analysis with the explicit
calculations of two-loop effects. For the $W$ exchange the two-loop
expression (\ref{2loopc}) reduces to the term with the anomalous
dimension $\gamma_{H}=-8$ which matches the result (\ref{Wcalc}).  In
the case of $a_\mu^{(Z)}$, inputting in Eq.\,(\ref{2loopc}) the initial
data (\ref{onelc}), (\ref{initVA}) and the mixings (\ref{genbVA}) we
observe at $\mu=m_\mu$ full agreement with the sum of
Eq.\,(\ref{open}) and (\ref{allclosed}).

The total two-loop result for $a_\mu^{(W)} +a_\mu^{(Z)}$ given in
Eq.\,(\ref{total2}) differs from that in Ref.~\cite{Degrassi:1998es}
in the term $(4/9) g^\mu_V g_V^fN_f Q_f$.  In \cite{Degrassi:1998es}
it is multiplied by a factor $(-Q_f)$.  This error originated in
Eq.\,(23) of \cite{Degrassi:1998es} for the two-loop mixing of the
operators $V_{\mu f}$ with $H$.  In accordance with the diagram $F_7$
in Fig.\,\ref{fig:fourmix}, the factor $Q_f^2$ in that equation should
be substituted with $Q_f Q_\mu \,$.

\subsection{The third loop effect}

To determine $h$ at the three-loop level from Eq.\,(\ref{RGeq1}) we have to
account for the running of $\alpha(\mu)$ and $c^{\Gamma;fg}(\mu)$ up
to the first loop,
\begin{eqnarray}
\alpha(\mu)&=& \alpha(M)\left[1+\frac{\alpha(M)}{2\pi}\,
\sum_f b_f \ln\frac{M}{\{\mu,m_f\}}\right] \,,\\[2mm]
\label{alpha1}
c^{\Gamma;fg}(\mu)&=&-\frac{\alpha(\mu)}{2\pi}\sum_{\Gamma',f',g'}
\gamma^{^{\Gamma; fg }}_{_{\Gamma'
;f'g'}}\,c^{\Gamma';f'g'}(M)\ln\frac{M}{\{\mu,m_{f'g'}\}}
\,.
\label{crun}
\end{eqnarray}
Using this expressions as well as the two-loop solution (\ref{2loopc})
for $h(\mu )$ we find,
\begin{eqnarray}
h^{(3)}(m_\mu)&=&\frac{\alpha^2(M)}{8\pi^2}\left\{\gamma_{_H} h(M)
\left[\gamma_{_H}\ln^2\frac{M}{m_\mu}+
\sum_f b_f L_f \right]\right.
\nonumber\\[1mm]
&&+\left.
\sum_{f,g, \Gamma}\beta_{\,\Gamma;fg}
\,c^{\Gamma;fg}(M)\left[\gamma_{_H}\ln^2\frac{M}{\{m_{fg},m_\mu\}}+\sum_l
b_l L_{l fg}
\right]
\right.\nonumber\\[1mm]
&&+\left.
\sum_{}\beta_{\,\Gamma;fg}\,
\gamma^{^{\Gamma; fg }}_{_{\Gamma'
;f'g'}}\,c^{\Gamma';f'g'}(M) L^{fg}_{f'g'}
\right\}.
\label{3loop}
\end{eqnarray}
Here 
\begin{eqnarray}
L_f&=&\ln^2\frac{M}{\{m_f,m_\mu\}}+2\theta(m_f-m_\mu)
\ln\frac{m_f}{m_\mu}\ln\frac{M}{m_f}\,,
\\[1mm]
L_{l fg}&=&\ln^2\!\frac{M}{\{ m_l, m_{fg},m_\mu\} }
+2\theta(m_l\!-\!\{m_{fg}, m_\mu\})\ln\frac{m_l}{\{m_{fg},m_\mu\}}
\ln\frac{M}{m_l}\,,
\nonumber\\[1mm]
L^{fg}_{f'g'}&=&\ln^2\!\frac{M}{\{ m_{fg},m_{f'g'},m_\mu\} }
+2\theta(m_{f'g'}\!-\!\{m_{fg},
m_\mu\})\ln\frac{m_{f'g'}}{\{m_{fg},m_\mu\}}
\ln\frac{M}{m_{f'g'}}\,.\nonumber
\end{eqnarray}

\begin{figure}[ht]
\begin{minipage}{16.cm}
\begin{tabular}{c@{\hspace{25mm}}c}
\psfig{figure=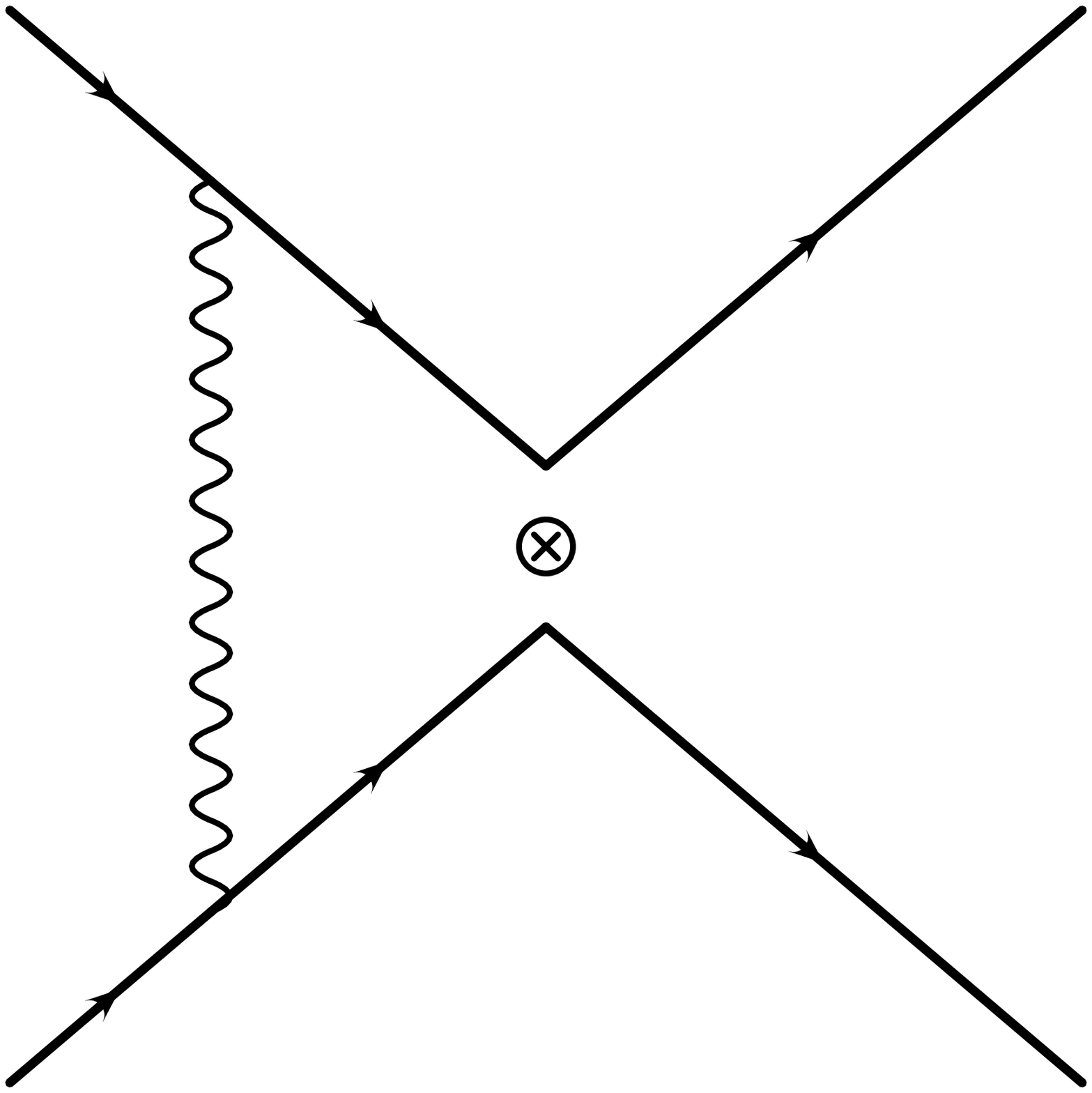,width=30mm}
&
\psfig{figure=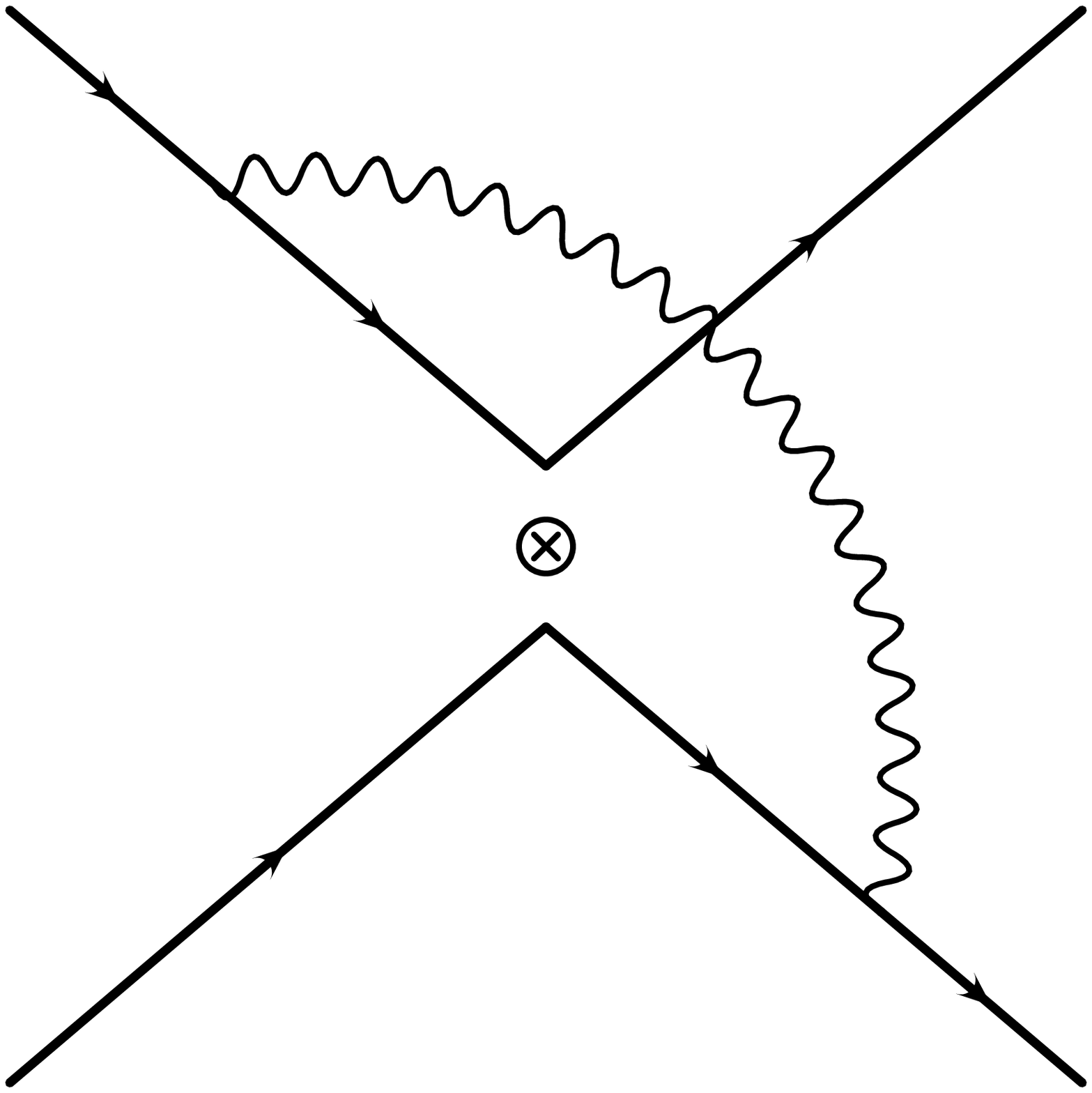,width=30mm}
\\[1mm]
 (a) & (b) 
\\[1mm]
\end{tabular}
\end{minipage}
\caption{\sf\small Renormalization of four-fermion operators: without 
annihilation.}
\label{fig:4ferm12}
\end{figure}

 Additional input, needed for the calculation of $c^{(3)}_H$, is the
anomalous dimension matrix for four-fermion operators.  The anomalous
dimension matrix  $\gamma^{{\Gamma'; fg }}_{{\Gamma
;f'g'}}$ is determined from the
one-loop diagrams in Fig.~\ref{fig:4ferm12} and Fig.~\ref{fig:4ferm34}.
The calculations are relatively straightforward: the diagram
Fig.\,\ref{fig:4ferm12}(b) produces no logs in the Landau gauge and can
be omitted, the diagram Fig.\,\ref{fig:4ferm34}(a) reduces to
Fig.\,\ref{fig:4ferm34}(b) by means of the Fierz transformations
(\ref{fierz}).  The result for nonvanishing entries in $\gamma^{{\Gamma';
fg }}_{{\Gamma;f'g'}}$ is
\begin{eqnarray}
\gamma^{{V; fg }}_{{A;f'g'}}&
=&-6Q_fQ_g \,\delta ^f_{f'} \delta ^g_{g'}
-\frac{2}{3}\,Q_f Q_{g'}\,\delta^{fg}\delta^f_{f'} 
-\frac{2}{3}\,Q_f Q_{f'}\,\delta^{fg}\delta^f_{g'} \,,\nonumber\\[1mm]
\gamma^{{A; fg }}_{{V;f'g'}}&=&-6Q_fQ_g \,
\delta ^f_{f'} \delta ^g_{g'}
\,,\nonumber\\[1mm]
\gamma^{{V; fg }}_{{V;f'g'}}&=&
-\frac{2}{3}\left[ N_fQ_fQ_{f'}\delta
^{g}_{g'} +N_f Q_fQ_{g'}\delta ^{g}_{f'}
+N_gQ_gQ_{g'}\delta ^{f}_{f'}\right.
\nonumber\\[1mm]
&&
\left.\quad+N_g Q_gQ_{f'}\delta ^{f}_{g'}
+Q_{f}Q_{g'}\delta^{fg}\delta^f_{f'}
+Q_{f}Q_{f'}\delta^{fg}\delta^f_{g'}\right]
\end{eqnarray}

Let us now use the expression (\ref{3loop}) to calculate the third
loop for the $W$ exchange. In this case the effective Lagrangian
(\ref{LW}) at $\mu=m_W$ contains only quark operators.  They do not
mix directly with $H$ so the second term in (\ref{3loop}) does not
contribute. In the last term it is the mixing of quark operators with
${\cal O}_{V;\, q \mu }$ which provides a nonvanishing
contribution. Another contribution comes from the first term in
(\ref{3loop}), associated with the anomalous dimension of $H$. Using
the initial data (\ref{initW}) for $c^{\Gamma;fg}(M)$ after some
simple algebra we arrive at 
\begin{eqnarray}
\label{eq105}
&&\!\!\!h^{(W)}(m_\mu)_{LL}
=h_{(W)}^{(1)}+h_{(W)}^{(2)}+h_{(W)}^{(3)}
=\frac{10}{3}-\frac{40}{3}\,\frac{\alpha(m_W)}
{\pi}\,\ln\frac{m_W}{m_\mu}\\[1mm]
&&+
\frac{\alpha^2}
{\pi^2}\left\{
\frac{10}{3}\left[8\ln^2\frac{m_W}{m_\mu} -
\!\sum_{f} b_f L_f(M=m_W)
\right]
\!-\frac{32}{81}\,\ln\frac{m_W}{m_Q}\ln\frac{m_W}{m_c}
-\frac{32}{81}\,\ln^2\frac{m_W}{m_Q} 
\right\},\nonumber
\end{eqnarray}
where we introduced $m_Q$ as an effective IR cutoff for the light
quark loops, $q=u,d,s$, implying $m_Q$ is larger then $m_\mu\,$.\footnote{%
Note that in sections \ref{sec:EW} and
\ref{sec:had} we used $m_s =0.5$ GeV different from $m_u=m_d=0.3$ GeV
but here we make the
simplification $m_u=m_d=m_s=m_Q$. That difference has no numerical
impact.}
We also added the first and second loop to make easier to follow a
numerical comparison. Taking $m_Q=0.3\,\mbox{GeV}$,
$m_c=1.5\,\mbox{GeV}$, $m_b=4.5\,\mbox{GeV}$ we get numerically,
\begin{eqnarray}
h^{(W)}(m_\mu)_{LL}&=& h_{(W)}^{(1)}\left[
1-26.5\,\frac{\alpha(m_W)} 
{\pi}+\frac{\alpha^2}{\pi^2}\,(352+359-6)\right]\nonumber\\[1mm]
&=&h_{(W)}^{(1)}\left[ \,1-0.067+0.0038\right]\,,
\label{eq106}
\end{eqnarray}
showing that the third loop effect is quite small. It is dominated
by the the anomalous dimension term $\propto \gamma_{H}^2$ (the first
number 352 in $\alpha ^2$ term) and by the cross term
$\propto\gamma_{H} b$ between the anomalous dimension and running of
$\alpha$ (the number 359), the four-fermion operators contribute very
little, ($-6$). Moreover, the effect of the third loop becomes even
smaller if we use $\alpha(m_\mu )$ instead of $\alpha(m_W )$ in the
second loop: this changes the $\gamma_{H} b$ term in the third
loop, $359 \to -252\,$.

Finally, we note that if we shift to the usual fine structure constant,
$\alpha=1/137.036$, via the full leading log relation
\begin{equation}
  \alpha(m_W) = \alpha + {2\alpha^2 \over 3\pi}
\sum_f N_f Q_f^2 \ln{m_W\over m_f} \simeq 1/129,
\end{equation}
it generates from the ${\cal O}(\alpha(m_W))$ terms in
Eqs.~(\ref{eq105}) and (\ref{eq106}) additional ${\cal O}(\alpha^2)$
contributions that amazingly cancel (numerically) the 0.0038 in
Eq.~(\ref{eq106}). So, in terms of the expansion parameter $\alpha$,
the leading log three-loop corrections are essentially zero. Such a
complete cancellation appears to be a numerical coincidence.

The algebra is more tedious in the case of the $Z$ exchange.  We set
$s_W^2=1/4$ which is a very good approximation numerically. It
simplifies the analytic result due to vanishing of the leptonic vector
couplings $g_V^{e,\mu,\tau}$ at this value of $s_W^2$, leaving us with
fewer operators.  We also combine the $W$ and $Z$ exchanges neglecting
the $m_W$, $m_Z$ mass difference --- again quite a good
approximation. (The difference will be in the three-loop NLL.)

We present the final result for the three-loop part of $a_\mu^{\rm
EW}$ in the form similar to that used in Ref.~\cite{Degrassi:1998es},
\begin{eqnarray}
a_\mu^{W, Z}(\mbox{3-loop})_{LL}
=\frac{G_\mu m_\mu^2}{8\pi^2\sqrt{2}}
\cdot\frac{5}{3}\,\frac{\alpha^2}{\pi ^2}\left(A_l +A_q +B_1+B_2\right),
\end{eqnarray} 
where $A_{l,q}$ come from lepton and quark terms in Eq.\,(\ref{3loop})
containing $\gamma_{H}^2$, $\beta \gamma $
\begin{eqnarray}
&&
\hspace*{-10mm}
A_l={2789\over 90}\ln^2{m_Z\over m_\mu}
-{302\over 45} \ln^2{m_Z \over m_\tau}
+{72\over 5}\ln {m_Z \over m_\tau}\ln {m_Z \over m_\mu}\,,
 \\[1mm]
&&
\hspace*{-10mm}
A_q = -{2662\over 1215}\ln^2{m_Z\over m_b}
+{11216\over 1215}\ln^2{m_Z\over m_c}
+{1964\over 405}\ln^2{m_Z\over m_Q}
+{24\over 5}\ln{m_Z\over m_b}\ln{m_Z\over m_\mu}
\nonumber \\[1mm]
&&
\hspace*{-4mm}
-{96\over 5}\ln{m_Z\over m_c}\ln{m_Z\over m_\mu}
-{48\over 5}\ln{m_Z\over m_Q}\ln{m_Z\over m_\mu}
+{32\over 405} \ln{m_Z\over m_b}\ln{m_Z\over m_c}
+{32\over 135} \ln{m_Z\over m_b}\ln{m_Z\over m_Q}\,,
\nonumber
\end{eqnarray}
and $B_{1,2}$ are due to the $\gamma _H b$, $\beta b$ terms involving
running of $\alpha\,$,
\newpage
\begin{eqnarray}
&&
\hspace*{-10mm}
B_1=-{179\over 45}\left({1\over 3} \ln^2{m_Z\over m_b}+\ln^2{m_Z\over
m_\tau }+{4\over 3} \ln^2{m_Z\over m_c}+2 \ln^2{m_Z\over m_Q}+2
\ln^2{m_Z\over m_\mu }\right)\nonumber\\[1mm]
&&
\hspace*{-4mm}
+{2\over 5}  \left(\ln^2{m_b\over m_\tau }+{4\over 3} \ln^2{m_b\over m_c}
+2 \ln^2{m_b\over m_Q}+2 \ln^2{m_b\over m_\mu }\right)
-{8\over 5}  \left(2 \ln^2{m_c\over m_Q}+2 \ln^2{m_c\over m_\mu }\right)
\nonumber\\[1mm]
&&
\hspace*{-4mm}
+{6\over 5}  
\left({4\over 3} \ln^2{m_\tau \over m_c}+2 \ln^2{m_\tau \over m_Q}
+2\ln^2{m_\tau \over m_\mu }\right)-{8\over 5}\ln^2{m_Q\over m_\mu }\,,
\nonumber\\[2mm]
&&
\hspace*{-10mm}
B_2={2\over 5}\left[2 \ln {m_Z \over m_\mu }+2 \ln {m_Z \over m_Q}
+{4\over 3} \ln {m_Z \over m_c}+ \ln {m_Z \over m_\tau }
+ {1\over 3}\ln {m_Z \over
m_b}\right] 
\nonumber\\[2mm]
&&
\times\left[{215\over 9}\ln {m_Z \over m_\mu }-4\ln {m_Z \over m_Q }
-8 \ln {m_Z \over m_c}+6 \ln {m_Z \over m_\tau  }+2\ln {m_Z \over m_b }
\right].
\end{eqnarray}
The $B_2$ term can be fully absorbed into the two-loop part of
$a_\mu^{\rm EW}$
given in Eq.\,(\ref{total2}) if $\alpha $ there is substituted by $\alpha
(m_\mu)$ instead of $\alpha(m_Z)\,$ which we used in our derivation
above.

Comparing with the results in Ref.~\cite{Degrassi:1998es} we see that
our $B_1$ coincides with their $B$ but the sum $A_l+A_q$ is somewhat
different from $A$ in \cite{Degrassi:1998es}. This is due to a few
reasons. One was already discussed: in Eq.\,(23) of
\cite{Degrassi:1998es} for the two-loop mixing of the operators
$V_{\mu f}$ with $H$ the factor $Q_f^2$ should be changed to $Q_f
Q_\mu\, $.  In addition, there is a difference in the one-loop
anomalous dimensions of four-fermion operators. In
\cite{Degrassi:1998es} an extra factor 2 is ascribed to the penguin
diagrams in Fig.\,\ref{fig:4ferm34}.  To correct this the factor $1/2$
should be introduced in Eqs.\,(33), (36), (38--41) of
\cite{Degrassi:1998es} and in Eq.\,(35) the $52/3$ should be changed
to $44/3$.

Numerically,
\begin{eqnarray}
A_l=1696\,,\qquad A_q=-507\,,\qquad B_1=-774\,,\qquad B_2=1916\,,
\end{eqnarray}
where we used the same values for the quark masses  as above.
Altogether, the three-loop correction is
\begin{eqnarray}
a_\mu^\mathrm{EW}(\mbox{3-loop}) = 
a_\mu^\mathrm{EW}(\mbox{1-loop}) \left({\alpha\over\pi}\right)^2
(A_l+A_q+B_1) \simeq 0.4\times 10^{-11}\,, 
\end{eqnarray}
where it is implied that $\alpha(m_\mu )$ is used for the two-loop
part. The numerical value is close to that given in
\cite{Degrassi:1998es}.  The reason for this final agreement is that
the $A$-part of the three-loop result is numerically dominated by
$\gamma _H^2$ and large mixing of axial operators with the dipole
operator, followed by the running of the dipole.  These pieces, as
well as the $B$-part, are correct in \cite{Degrassi:1998es}. If
$\alpha(m_Z)$ is used in the two-loop part the third loop is somewhat
larger, $a_\mu^\mathrm{EW}(\mbox{3-loop}) \simeq 2.4\times 10^{-11}$.

Of course, in both cases one must reevaluate $a_\mu^{\rm EW}$(2-loop) with
the shifted coupling at scale $m_\mu$ or $m_Z$.  In that way, scale
insensitivity is restored.  In addition, because the effective
couplings are larger than the usual fine structure constant, $\alpha$,
a transition to this $\alpha$ in the two-loop part of $a_\mu^{\rm EW}$
induces additional negative contributions.
  Remarkably, those negative contributions cancel with the
above explicit three-loop results to about $0.1\times 10^{-11}$.  (The
cancellation is similar and of course related to the even more
complete cancellation pointed out for the $W$ contribution alone.)
Hence, to a good approximation, the leading-log higher order
contribution is zero or at least negligible.

\newpage
\section{Summary}
\label{sec:five}

Having addressed a variety of computational issues, 
\begin{itemize}
\item
Small, previously neglected, 2-loop contributions suppressed by
factors of $(1-4s_W^2)$ that come from $\gamma$-$Z$ mixing and the
renormalization of $\theta_W$,
\item
Strong interaction modifications of quark loop diagrams, and
\item Leading log 3-loop effects,
\end{itemize}
we are now in a
position to update the Standard Model prediction for $a_\mu^{\rm EW}$
and assess its degree of uncertainty.  

Small effects due to $\gamma$-$Z$ mixing and our choice of  $\theta_W$
renormalization have now been included in Eqs.\,(\ref{eq17}) and
(\ref{eq18}).  Because of the $(1-4s_W^2)$ suppression factor, their
total impact is rather small, shifting the value of $a_\mu^{\rm EW}$
down by about $0.4\times 10^{-11}$.  

More important are strong interaction effects on the quark triangle
diagrams in Fig.\,\ref{fig:triangle}, particularly in the case of light
quarks.  It was shown that short distance contributions are unmodified
(thereby, hopefully, eliminating controversy in the literature).
However, QCD can affect their long-distance properties.  In the case
of the first generation of fermions a detailed operator product
expansion analysis and effective field theory calculation led to a
shift relative to the free quark calculation (with constituent quark
mass) by
\begin{eqnarray}
\Delta a_\mu^{\rm EW}[e,u,d]_{\rm QCD}-\Delta a_\mu^{\rm
  EW}[e,u,d]_{\rm free~quarks} = +1.7\times 10^{-11}.
\end{eqnarray}
For the second generation, comparison of the free quark calculation
with the more precise evaluation in Eq.\,(\ref{secgenN}) shows no
significant numerical difference.  However, the more refined analysis
now indicates very little theoretical uncertainty.  So, the total
hadronic uncertainties in $a_\mu^{\rm EW}$ would seem to be well
covered by an uncertainty of $\pm 1\times 10^{-11}$ or even less.

Finally, after a detailed renormalization group analysis, the leading
log three-loop contributions turned out to be extremely small.  In fact,
they are consistent with zero, to our level of accuracy $\sim 0.1
\times 10^{-11}$, due to a remarkable cancellation between anomalous
dimensions and running coupling effects.  Uncalculated three-loop NLL
contributions are expected to be of order
\begin{eqnarray}
\frac{G_\mu m_\mu^2}{8\sqrt{2}\pi^2} \left({\alpha \over \pi }
\right)^2
\ln{m_Z^2 \over m_\mu^2} \simeq 8\times 10^{-14},
\end{eqnarray}
which is negligible unless enhanced by an enormous factor.
Nevertheless, we assign
an overall uncertainty of $\pm 0.2\times 10^{-11}$ 
to $a_\mu^{\rm EW}$ for uncalculated three-loop NLL
contributions.

So, in total we find a small shift in $a_\mu^{\rm EW}$ (for $m_H
\simeq 150 $ GeV) from the previously quoted value of $152(4) \times
10^{-11}$ to a slightly larger (but consistent) value
\begin{eqnarray}
 a_\mu^{\rm EW} = 154(1)(2) \times 10^{-11}
\end{eqnarray}
where the first error corresponds to hadronic loop uncertainties and
the second to an allowed Higgs mass range of 114 GeV $\alt m_H \alt$
250 GeV, the current top mass uncertainty and unknown three-loop
effects.

\begin{acknowledgments} 
Authors are grateful to S.~Adler, S.~Eidelman, G.~Gabadadze,
E.~D'Hoker, E.~Jankowski, Mingxing Luo, K.~Melnikov, M.~Shifman,
M.~Voloshin and A.~Yelkhovsky for helpful discussions. A.V. is
thankful for the hospitality to the Aspen Center for Physics where a part 
of this work was done. We also appreciate communications with
G.~Degrassi, G.~F.~Giudice,
M.~Knecht, S.~Peris, M.~Perrottet and E.~de Rafael.

This research was supported in part by the Natural Sciences and
Engineering Research Council of Canada, the DOE grants
DE-AC03-76SF00515 and DE-FG02-94ER408, NATO Collaborative Linkage
Grant, and the Alexander von Humboldt Foundation.

\bigskip

\noindent
{\bf Note added} (May 2006):\\[2mm]
In regard to  the OPE analysis discrepancies with Ref.\ \cite{Knecht:2002hr} discussed in Sec.\ III.E.3
we would like to note that the authors  of Ref.\ \cite{Knecht:2002hr} in their later publication
[M.\ Knecht, S.\ Peris, M.\ Perrottet, and E.\ de Rafael, JHEP {\bf 0403}, 035 (2004)]
expressed their acceptance of our analysis. We also would like to acknowledge their comment
in this publication in regard to our discussion (after Eq.\ (46))
of the operator $\tilde F^{\alpha\beta}{\rm Tr}\, G_{\mu\nu}G^{\mu\nu}$.
Although we did not account for its contribution anyway they correctly emphasized 
that this operator does not appear in the operator product expansion 
because of its chiral features.

 We thank Jens Erler for pointing out a mistake in
Eq.\ (55) in the earlier version of this paper.

\end{acknowledgments}

\end{document}